# The fair reward problem:
# the illusion of success and how to solve it

*Didier Sornette, Spencer Wheatley and Peter Cauwels*
ETH Zurich

**Abstract:** Humanity has been fascinated by the pursuit of fortune since time immemorial, and many successful outcomes benefit from strokes of luck. But success is subject to complexity, uncertainty, and change – and at times becoming increasingly unequally distributed. This leads to tension and confusion over to what extent people actually get what they deserve (i.e., fairness/meritocracy). Moreover, in many fields, humans are over-confident and pervasively confuse luck for skill (I win, it's skill; I lose, it's bad luck). In some fields, there is too much risk-taking; in others, not enough. Where success derives in large part from luck – and especially where bailouts skew the incentives (heads, I win; tails, *you* lose) – it follows that luck is rewarded too much. This incentivizes a culture of gambling, while downplaying the importance of productive effort. And, short term success is often rewarded, irrespective, and potentially at the detriment, of the long-term system fitness. However, much success is truly meritocratic, and the problem is to discern and reward based on merit. We call this *the fair reward problem*. To address this, we propose three different measures to assess merit: (i) *raw outcome*; (ii) *risk-adjusted outcome,* and (iii) *prospective*. We emphasize the need, in many cases, for the deductive *prospective approach*, which considers the potential of a system to adapt and mutate in novel futures. This is formalized within an *evolutionary system*, comprised of five processes, inter alia handling the *exploration-exploitation trade-off*. Several human endeavors – including finance, politics, and science – are analyzed through these lenses, and concrete solutions are proposed to support a prosperous and meritocratic society.

## 1. Introduction: 'O Fortuna'

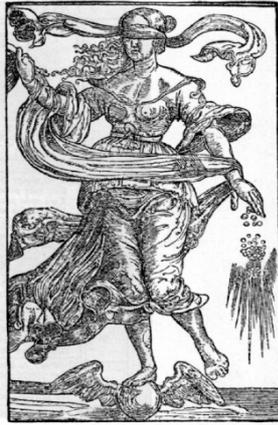

*Figure 1. Lady luck, blinded. Source unknown.*

We like to think that our good fortune is a result of our talent, hard work, willpower, etc. In other words that we live in a *meritocratic* society. In fact, decades of modern research in many disciplines[1] show that luck often plays an important role, and disentangling the relative contributions of skill and luck in success is hard. Select findings have been made accessible to a broader public in best-selling books[2]. Naturally, humanity has always tried to make sense of fortune, and benefit from 'luck'. We profit from an ancient enlightened view[3]: In Ancient Greece, their 'theory' of luck was *Tyche*: the goddess of luck, fortune, and fate. The Romans called her *Fortuna*. In medieval art, she was depicted as carrying a cornucopia (prosperity), a ship's rudder, and operating the *Rota Fortunae (wheel of fortune)* – symbolizing eternal *cyclical dynamics* between prosperity and crisis. At times, she could be *blinded by a veil (luck)*, but at times rewarding the virtuous, and equally, punishing the unvirtuous *(merit)*. Worship and blame of her would tend to spike around periods of upheaval and great uncertainty.

Seneca's *Agamemnon* (55 AD), in its chorus, addresses *Fortuna*, with a powerful message on the multiple forces that conspire against great orders, bringing them to great ruin[4]:

*O Fortuna, who dost bestow the throne's high boon with mocking hand, in dangerous and doubtful state thou settest the too exalted. Never have sceptres obtained calm peace or certain tenure; care on care weighs them down, and ever do fresh storms vex their souls.*

*What palace has not crime answering crime [elite power struggles] … ? … Law, shame, the sacred bonds of marriage, all flee from courts. Hard in pursuit comes grim Bellona [goddess of war] … and Erinys [goddess of vengeance], … which any hour brings low from high estate.*

*Though arms be idle and treachery give o'er, great kingdoms sink of their own weight, and … 'tis the high hills that the lightnings strike; large bodies are more to disease exposed …*

*Whatever Fortune has raised on high, she lifts but to bring low. Modest estate has longer life; then happy he whoe'er, content with the common lot, with safe breeze hugs the shore…*

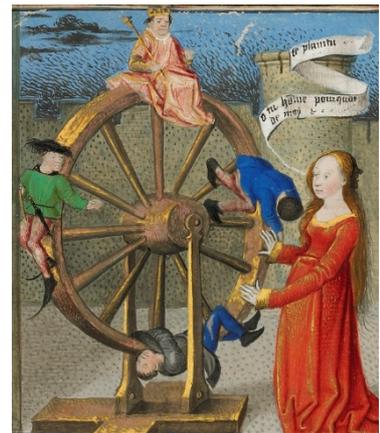

*Figure 2. Lady luck and her rota fortunae. "Coëtivy Master" (Henri de Vulcop, 1450 - 1485) The J. Paul Getty Museum, Los Angeles, Ms. 42 (91.MS.11).*

We gladly take the *baton* and continue in this tradition: As a definition, one is *lucky* if, in matters of chance, one tends to obtain fortunate outcomes. Bad luck exists as the opposite. Relating to chance, luck hence depends on what is defined as random – being an epistemic concept, e.g., determined by the limits/boundaries of ones' system model, beyond which probabilistic descriptions must be employed. Fortuna (in epistemic form) lives beyond the boundary, and (in stochastic form) within 'black boxes' inside of the system. Reflecting on *the Agamemnon model*, a clear motif is: that which is elevated by luck will also fall by luck, lacking the merit to sustain it. This self-correcting theory is impressive[5]. It is of high interest to consider its efficiency in potentially singular modern times, and how it can be adapted and retrofitted. Indeed, how do the performance of these mechanisms function in today's modern technological welfare states?

---

[1] Complex systems, cognitive sciences, behavioral economics, political sciences, finance, business, management, sports, gambling... There have been many exceptional and original contributions to the academic literature (see for example early works of Merton, 1961, Langer, 1975, Montroll and Schlesinger, 1982 and Dorner et al, 1990).
[2] e.g., Taleb, 2005; Mauboussin, 2012; Tetlock and Gardner, 2015; Kahneman, 2011; Frank, 2016.
[3] Wikipedia.org keywords: Tyche, Fortuna, Rota Fortunae.
[4] Translation by Frank Justus Miller (https://www.theoi.com/Text/SenecaAgamemnon.html)

An instructive starting point is the 'birth lottery' – from family wealth to genetics[6] to upbringing– means that practically all determinants of success are largely subject of luck, with effects decaying slightly over time as individual will-power takes over. To embed more broadly, consider the so-called 'frozen accidents' of Nobel laureate Murray Gell-Mann (Gell-Mann, 1994), who argues that the universe and the nature of life are the result of chance events and the effective complexity of the world receives only a small contribution from the fundamental laws. However, e.g., extending the individual to the immediate family, it becomes clear that much of the luck of the child is the merit of the parent. Further, it is natural that both merit and (productive) risk-taking – e.g., exploration rather than pure gambling – must be rewarded. Dealing with the *fair reward problem*, relating to social systems, a definition of luck that finds a middle-ground between these opposing realities must be found; In societal terms, meritocracy and productive incentives are clearly a force for good (on the free will side). But equity/equality issues (on the "luck"/fatalism side) cannot be ignored, as we strive to optimize overall prosperity and sustainability.

How skill and luck contribute, each separately, to success depends on the game. Excluding inherent stochasticity, in a regular environment, with clear and simple rules, and which provides sufficient time and opportunity to learn through trial-and-error, merit will be the principal contributor to success. In the opposite case, luck will tend to dominate. A univariate spectrum mapping the importance of luck in a number of *games* was given by Mauboussin (2012) as visualized in Figure 3. Here, for each activity, its position on the spectrum is driven by the variance of the actual observations/outcomes relative to the variance of the results that would be seen if the game were purely run by chance (e.g., if a soccer game were driven by coin tosses). Naturally, chess and roulette are on opposite ends of the spectrum. This approach is very interesting, but limited[7].

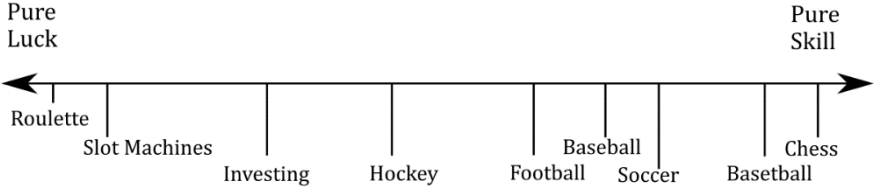

*Figure 3: Different activities in sports, investments and gambling, ranked on the skill-versus-luck continuum (adapted from Mauboussin, 2012).*

For activities on the right side, rewarding success (i.e., outcome) is adequate, since merit and success are highly correlated. And, in real-world endeavors that fall to the left side, success and failure are both largely driven by luck. Insofar as societies reward success but limit the downside of failure, risk-taking may over-develop: incentivizing gambling and deterring genuine efforts. Where true, is this good for society? How can we improve? Here we open 'Pandora's box' and challenge the current model.

Taking an approach that is novel in its comprehensiveness, we contribute the following. First, we provide a *static* framing of the (fair) *reward problem* via the *skill-success gap*. Next, to better solve the *reward problem*, we emphasize the need to consider *dynamic* games, affected by external (e.g., environmental) and internal (e.g., ethos/cultural paradigm) factors. We formalize the game structure, and construct a conceptual *Greek operating system* to navigate it. Further, depending on system observability, we introduce and characterize three measures for the merit/performance of a system/process: raw outcome, risk adjusted, and prospective. We conclude by reflecting on a range of important fields and problems, in view of the advances made within.

---

[5] This will resonate with the characteristic time discussed below (see section 5.2, equation (4), where time is the differentiator between luck and skill through mean reversion.
[6] "The selfish gene" book by Richard Dawkins (2016) makes this point forcefully, by taking a gene-centric view of evolution, in which individuals of an ancestry-descendent genealogy are the connected successive "vehicles" of evolution of gene replicators.
[7] It depends on the competitiveness of a league (in super-elite competition, luck can become more decisive) and resulting intransitivity cycles with luck.

2. Framing the *reward problem*: the *skill-success gap*

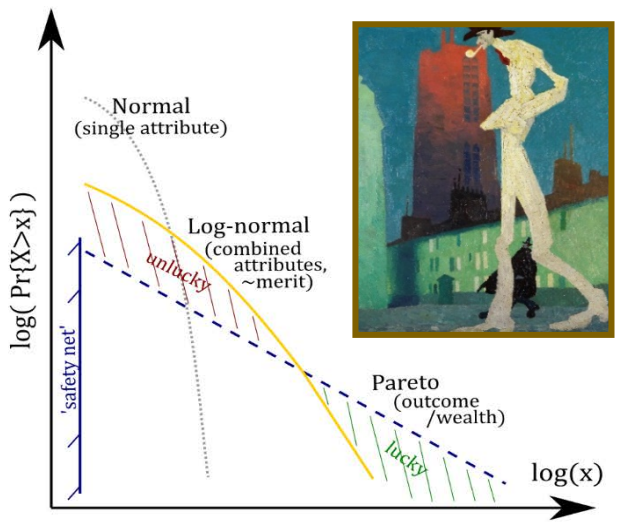

*Figure 4. The skill-success gap. The tails of the Normal, Log-normal, and Pareto distributions are given in double logarithmic scale. The gaps between the Log-normal, as a model for productivity or merit, measures the 'skill-success gap', comprising lucky and unlucky groups. Inset is "The White Man", by Lionel Feininger (1907) of the Carmen Thyssen-Bornemisza Collection (Inv. No. CTB.1972.15)*

Here, we provide a fundamental static framing of the *(fair) reward problem*: As the reader is likely well aware, even in a *relatively homogeneous* population, there are large deviations in wealth – which we reservedly take as proxy[8] for success – with the distribution being well approximated by the Pareto/power-law. On the other hand, in terms of physiological parameters/"phenotypes", are we not rather similar? How can this apparent gap between the mild distribution of human traits, and the wild distribution of wealth outcomes be explained? This was artfully captured in expressivist style by Feininger, where one man towers over the other (figure 4). Many blame luck, unequal opportunity, or corruption. On the other hand, others justify inequality as the direct and unavoidable consequence of hard work and productive risk taking. As with many polarizing political issues, the truth lies somewhere in between, and requires careful thinking to access.

Concretely here the wealth and single attribute distributions are observable, but the merit distribution lies between but is not directly observable. The distance between the wealth and merit curves defines the *skill-success gap*. Considering the ideal case of an equal opportunity meritocracy – i.e., no corruption and group discrimination – is it possible for heavy-tailed wealth distributions to emerge? Unequivocally the answer is yes. Here, we present a merit-based mechanism for this, and luck-based mechanisms follow in the subsequent section: While studying scientific output in research laboratories, the 1956 Nobel Prize winner in Physics, William Shockley, found very large spreads in productivity, with deviations up to a factor of one hundred between extreme individuals (Shockley, 1957). This was surprising as most human attributes and activities vary over much narrower limits: heart rates, physical performance (e.g., top running speed) or intellectual performance (e.g., school exam results, or IQ tests) seldom exhibit variability beyond a factor of 2-3 of the mean. The explanation, according to Shockley, is that *large changes in rate of production may be explained in terms of much smaller changes in certain attributes*. Suppose that an agent wants to finish a certain task in a given period of time, and that the outcome results from the combination of n factors (or sub-tasks) $F_1$, $F_2$, …, $F_n$. Quantitatively, the probability of completing this task can be approximated by the product of these factors. Then, the productivity (P) of this agent can be given by the following formula:

---

[8] Wealth is the current dominant measure of success. Of course, human prosperity is more encompassing, but introduces some subjective uncertainty. Wealth is conveniently and objectively observed. However, quoting William Bruce Cameron "not everything that counts can be counted, and not everything that can be counted counts". Goodhart's Law is also relevant: "When a measure becomes a target, it ceases to be a good measure", i.e., once a proxy metric is selected and then optimized, the optimization will not converge to the intended target. This is a recognized fundamental issue, e.g., leading to criticism of use of gross domestic product measures.

$$P \cong F_1 F_2 F_3 \dots F_n \qquad (1)$$

Thus, if one individual exceeds another by 50% in each of say 10 sub-tasks, their final productivity will be larger by a factor close to 58. In support of his theory, by statistical study of the rates of publications in research laboratories, Shockley found out that not simply the rates of observations, but rather its logarithm, had a normal distribution. The existence of this Log-normal distribution naturally follows[9] from equation (1).

These days, Shockley's theory increasingly makes sense, as automation leaves behind, as well as generates, mostly complex tasks -- requiring multiple productive attributes[10]. Indeed, when considering productivity, technology plays an amplifying transformational function. This is compounded by the apparent reality that, at some point, vanishingly few cognitive elite will be able to master the rapidly advancing technology. In the extreme, unlike a continuous normal distribution, productivity in these cases will be 'zero' for most, and huge for the few that can exploit the massive efficiencies and scalabilities inherent in software, automation, and so on. Put directly, there will probably be a hard cutoff at a certain intelligence level. In popular culture, the viciousness of this has been captured by the meme *learn to code*, hurled at the *have nots*, who lack computer programming abilities.

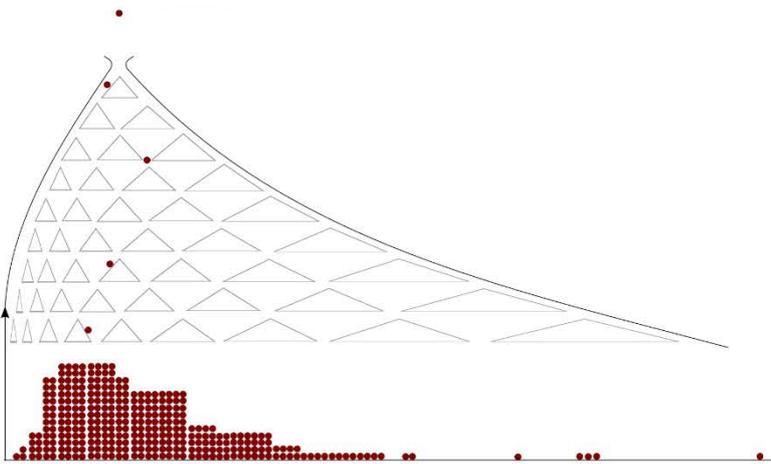

*Figure 5. 'Stahel board': Right-skewed adaptation of the Galton board to produce Log-normal variates rather than Normal variates, which would result from the case of a symmetric board with identical triangles. Reproduced from Limpert, Stahel & Abbt, 2001.*

To reinforce, statistical analysis of distributions across the sciences concluded that "*life is log-normal*"; without a single case where the Normal outperformed the Log-normal in terms of goodness of fit -- evidencing pervasive right-skew (Eckhard, Stahel, & Abbt, 2001). Indeed, both the Normal and Log-normal arise from a variety of forces working together: for the Normal they are additive, and for the Log-normal, multiplicative (visualized in figure 5). The hypothesis of Ekhard et al. rings true here: Much of (re)production in nature is driven my multiplication. E.g., the speed of production by chemical reaction is driven by the multiplication of the concentrations of the ingredients. The terms product and multiplication are even etymologically linked. And, the more complex the process, the more multiplicative levels.

It is hence surprising that many have modelled merit and related quantities by Normal distributions (Stewart, 1983; Crocker and Algina, 1986; Mauboussin, 2012; Pluchino et al., 2018). Especially in the realm of complex tasks and total output, such assumptions will tend to overestimate the gap between merit and outcome, as shown in figure 4. Mathematically, it is interesting to note that, when the standard deviation of the Log-normal is small, it is close to the Normal. However, the more factors that are included in (1) and the larger their (potentially heterogeneous) variances, the heavier-tailed the productivity distribution will be. In fact, with sufficiently high variance, the Log-normal can approximate a Pareto distribution – in that extreme, explaining all wealth inequality by actual variation in productivity! The log-normal thus provides a useful framework, encompassing edge cases of both massive and negligible gaps between productivity and wealth. It seems that the first edge case of *pure luck* can now be excluded.

---

[9] The logarithm of the product is the sum of the logarithms of the different factors. If the factors are independent, then -- to a good approximation, and if the central limit theorem is applicable -- their sum will be normally distributed, and hence so will be the logarithm of the productivity. Convergence to the Log-normal can be slow (see e.g. Redner, 1990).. But, if the individual factors are themselves Log-normal, then the product is immediately Log-normal.

[10] Talent, perseverance, skill, analytical intelligence, communication abilities, charisma, emotional intelligence, social skills, etc.

Another important factor of the skill-success gap lies in the ability to learn for experience (Yin et al., 2019), which requires embedding the occurrence of success in a more dynamical description based on attributes of tenacity/morale catalyzing the pursuit of success and the willingness to adapt/transform when needed.

In the next section, we exclude the *pure merit* edge case by describing important mechanisms that inject large amounts of luck into the success process.

### 3. Mechanisms of luck: the modern *rota fortunae*

Here, a non-exhaustive list of important yet under/mis-understood mechanisms that inject luck into success are described. More specific examples are in the final discussion.

### 3.1 Gibrat's law, proportional growth and the Matthew effect

The mechanism of proportional growth was first formulated by Yule (1925) to explain the distribution of species among genera of plants, which had been shown empirically by Willis to satisfy a power law distribution. Gibrat (1931) found that the rate of growth of a firm, a city or the wealth of a family is independent of its absolute size, hence the term proportional growth. Simon (1955) generalized to stochastic proportional growth to explain the power law distributions of word frequencies in documents, of numbers of papers published by scientists, of city population sizes, incomes, and species among genera. Merton (1968) documented the so-called *Matthew effect* in scientific reputation, where "*For to everyone who has more will be given, and he will have abundance; but from him who has not, even what he has will be taken away*" (Matthew 25:29), another incarnation of the proportional growth mechanism. Starting with Simon (1955), many authors have documented and derived that the mechanism of proportional growth leads generically (when combined with other minor ingredients) to the concentration of success often characterized by power law distributions (see for reviews in different domains: Ijiri and Simo, 1977; Mitzenmacher, 2003; Sornette, 2006; Saichev et al., 2009; Malevergne et al., 2013). The power law distribution results from the cumulative effect of stochastic (lucky) proportional growth terms, which goes already a long way towards explaining pervasive inequality so that a small fraction of the population controls a large fraction of the resources. See quantitative verifications of this in growing social networks (Zhang and Sornette, 2011), competing electronic products (Hisano et al., 2011) and crypto-currency market shares (Wu et al., 2018).

### 3.2 Winner-takes-all: a pervasive mechanism for luck in success

We live in a strongly connected society where ideas, new technologies, fashion, music, and so on, are adopted through viral interaction in social networks. Imitation and social influence are the key elements in a positive feedback mechanism that creates an effect of cumulative advantage for winners. As explained by Brian Arthur (1989)[11]: when economic objects compete, under increasing returns, insignificant events or minor differences in the initial conditions may be decisive – locking in the initial advantage. Through the feedback mechanism of adoption and improvement, the agent that by chance gains a lead, may tend to corner the market. In the limit, this becomes a winner-take-all market (WTAM) where the success is unrelated to the merit of the different economic agents. There are likely many luck-dominated fields largely driven by WTAM[12]. However, history shows us that also latecomers can overcome this strong path dependency, and gain a dominant or quasi-monopolistic position. Google, for example, launched in 1997 as a latecomer in the search engine ecology. But in less than three years, it *became the most popular search engine* (Barabási, 2014). Hence superior *fitness* (technology) can overcome stochastic influences.

---

[11] Arthur concludes: To the degree that the technological development of the economy depends upon small events beneath the resolution of an observer's model, it may become impossible to predict market shares with any degree of certainty
[12] In financial planning and budgeting, economic forecasting, marketing, design, fashion, research and development, trading, investing, mergers and acquisitions, private equity, venture capital, music, film, art and so on.

More insight into the nuances of these dynamics are available from so called *Tullock contests* (Tullock, 1980): In a typical situation, players that expend resources have a higher probability of winning, and have a total payoff depending on the winning prize, own effort, and the effort of rivals (Chowdhury and Sheremeta, 2010). By changing the competitiveness parameter, cost of winning, and spillover from losing, one can move from random allocation to a winner-takes-all mode. Importantly, small parameter modifications may lead to substantially different optimal effort levels and changes in the role of chance. E.g., when the prize does not justify the efforts of competition, the players let chance decide the outcome. Also, if the positive externality when losing increases relative to when winning, then players will compete less, reminiscent of *R&D contests* where property rights are not well protected.

3.3 Adverse selection

In any communication between individuals, there is a degree of asymmetry, where one party has more or better information than another. When there are conflicting interests, or when incentives are not aligned, false signals may be disseminated. The origin, the functioning and the consequences of this *dishonest signaling* (i.e., bluffing) is an established research subject in evolutionary biology.

In economics, the mechanism was developed by George Akerlof (1970) who proposed a thought experiment based on a second-hand car market with some 'lemons' (second hand cars of inferior quality): When sellers and buyers have asymmetric information about the quality of a good, buyers will automatically correct their bid price based on an estimation of the fraction and value of lemons in the market. This bid price will be somewhere in between the value of a good car and a lemon. Sellers, on the other hand, having superior information, will set their ask price higher than the bid price in case of a good car, and lower than the bid price for lemons. As a consequence, sellers of high-quality goods will exit from the market leading to an adverse selection of low-quality goods – potentially leading to malfunctioning of markets. This also applies to job-markets, where education is a signal of quality (Spence, 1973), and strategies exist to handle lack of/asymmetric information in such markets (Rothschild and Stiglitz, 1976). Finally, concerning financial markets, the *Grossman-Stiglitz paradox* states: if a market were informationally efficient, so that all information is reflected in market prices, there would be no incentive to acquire the information on which these prices are based (Grossman and Stiglitz, 1980)[13]. Hence near efficient markets are repelled from the point of full efficiency.

These works stress the difficulty of the reward problem, not only due to stochasticity, but further aggravated by asymmetric information in the signaling of success. These differences can drive out good quality and lead to adverse selection. Further, randomness in reward may also arise from the reward mechanism itself – i.e., due to arbitrariness, partial views (asymmetric information), or uncertainty of performance. Indeed, *group-level scoring* is also relevant, which tends to water down individually relevant information, leading to the potential destruction of cooperation in response to inequitable rewards (Duca and Nax, 2018): as freeloaders thrive, merit evaporates and chance mushrooms. Hence, providing means to measure and discern merit/skill should help. A mechanism for reward distribution in teams is discussed in 5.3.

3.4 Male-male competition and evolution

All living individuals share common ancestors, observable by genetic methods. The Most Recent Common Ancestor can be identified as the human being from which all others directly descended. The age of this genetic singleton is called the Time to the Most Recent Common Ancestor (TMRCA). Studies have shown that the matrilinear TMRCA is roughly twice that of the patrilinear.[14] In other words, females have propagated their genes across larger spans of time than males, who have been faster "forgotten" in the battle of genetic transmission.

---

[13] Akerlof, Spence and Stiglitz received the Nobel prize in economics for their work on asymmetric information.
[14] Based on these findings 'Eve' should have lived 170-240 thousand years ago, whereas 'Adam' would have wandered paradise's pastures 46-110 thousand years ago (Baumeister, 2010).

Using Agent Based Models, Favre and Sornette (2012) demonstrated that the strong unequal biological costs of reproduction between the two genders is the likely underlying explanation of unequal TMRCAs. Females -- having both scarce and costly reproduction potential -- engage in choosy selection, resulting in strong male-male competition and high risk-taking behaviors, with each male striving to signal as 'alpha'. Over human history most males failed, so that the trace of their genes faded away much more rapidly than those of females, explaining the factor 2 in the ages of the TMRCAs (Baumeister, 2010; Favre and Sornette, 2012). Importantly, the high-risk taking propensity of males, especially young males in the early years of reproduction, has survived in modern humans. Thus, even in the absence of problematic incentives, men have evolved to take much more risk than females.

Gender competitive-related differences are of particular importance in finance. Coates et al. (2010) document that success in financial markets increases the testosterone level in young male traders. As a consequence, more gender diversity amongst traders, as well as risk managers should have a positive effect on the stability of financial markets. Favre and Sornette (2012) conclude: "*…These considerations raise the question of how to adjust our cultures and/or design society rules in order to match better the evolutionary-based inherited human traits described above and the requirements of the modern ages.*"

Given such heterogeneities in human nature, clearly it is important to measure risk and account for the role of risk and hence chance events in the realization of outcomes.

<u>3.5 Heads, I win (it's skill). Tails, *you* lose (it's bad luck)</u>

Though people are fully aware of the concept of chance in a game, they still behave as if they are in control. Indeed, throwing dice, players behave as if they were controlling the outcomes -- being careful to throw softly if they want low numbers, throwing hard for high numbers. Ellen Langer referred to this phenomenon as *the Illusion of Control*[15] (Langer, 1975). She concluded that the illusion of control arises from the fact that people want to be in charge of their environment and avoid the anxiety from loss of control -- including over chance events, where people are convinced that they have the ability to beat the odds.[16]

Using computer simulated environments, Dietrich Dorner went beyond chance situations and investigated how humans make decisions in complex fields of reality. His scenarios included non-linear dynamics, network dependencies, time-delays and oscillations resulting from feedback processes. In a paper entitled the Logic of Failure, Dorner et al. (1990) explain that people do not want to be confronted with the consequences of their actions. In that way, they can maintain an *illusion of competence*. Dorner et al. call this acting *ballistically*. Without self-reflecting examination and critique, individuals delude themselves in having solved problems merely by means of their action without any need to look back and check for the consequences and the effectiveness of their deeds. Examining success and failure, most people hardly look beyond the superficial, immediate, and obvious. As such, and by human nature, they attribute desirable outcomes to internal factors, but blame externalities and bad luck for failures. They are hard-wired to over-attribute success to skill and to underestimate the role of chance. Langer refers to this as the "just world" view.

It is hence important to design incentive and reward systems with such cognitive biases in mind. Clearly ballistic and illusion of control thinking coupled with rewards based on short term outcomes can be at the detriment of long term system fitness, leading to excessive risk taking. Another important aspect of many social systems are safety nets and bailouts, which provide a lower bound on exposure to negative outcomes for those that fall on hard times – by luck or by choice. In some cases ('too big to fail' comes to mind), this asymmetry in exposure to consequences can also encourage unhealthy risk-taking. Coupling this with the asymmetric attribution of merit and luck, depending on the outcome, is a strong dynamic in the award of bailouts. For instance, concerning financial and economic crises, 'surely the fault cannot lie with the financial masters of the universe'. However when markets go up, "those bonuses are well and fully deserved". Ballistic and illusion of control thinking together with possibility of bailouts hence forms a corrupting feedback loop, whereby public bailouts tend to be given to those that do not deserve them, perpetuating a nepotistic (anti-meritocratic) system. This combination is summarized in the title of this sub-section.

---

[15] An expectancy of a personal success probability inappropriately higher than the objective probability would warrant.
[16] The greatest satisfaction/feeling of competence results from controlling the seemingly uncontrollable (e.g., the game golf). As a former golfer, one of the authors can confirm that it is all about the satisfaction of controlling that tiny white ball at high speed – requiring great power -- across long distances, and putting it into a little hole at the end.

3.6 The big statistics *replication crisis*

The technology and practice of statistics lie both at the heart of the issues discussed here – separating signals (skill) from noise (luck) – as well as in the fundamental advance of knowledge, and related decisions under uncertainty – in and out of the academy. Hence, statistical and deterministic sciences together have come to explain much of what was formerly epistemic uncertainty; without them, our model of the world would leave far more to chance. Since quality and success rely heavily upon *decision-making* (including design, and process control) – itself relying on our collective *knowledge base* to know pros, cons, and uncertainties – sound statistical reasoning is a strong determinant of the quality/merit of decisions.

As with all technologies, intentionally or not, their use can be both beneficial and harmful. For instance, consider if our *knowledge base* – e.g., propositions accepted through statistical significance testing – were to become corrupted by falsehoods ('fake news'), and hence cast into doubt? Fatalistic views could grow, reducing beneficial efforts and actions. Also, actions could be taken on the basis of spurious propositions. This would harm performance, weaken meritocracy, and inject more chance into everyday life. What is the risk and extent of such a development?

Use of statistical decisions comes with intrinsic risk: type 1 and type 2 errors, also called *false alarm* and *miss*. To immediately impress the *gravity and necessity* of this: it is the inevitable balancing of the risks (and benefits) of convicting an innocent versus letting a guilty criminal go free. In each instance an innocent is put on trial, the court is playing Russian Roulette with their freedom. But, without this, we cannot decide and we cannot have justice. The doctor cannot tell the patient if they are sick or not. The policy-maker cannot decide if measured climate change is worthy of action. Technically, in a *hypothesis test,* one specifies the type 1 error probability ('*a*') as the p-value decision threshold: e.g., 'reject null hypothesis if p-value < a'. By definition, if the null hypothesis of 'no effect' is true, *a* is the probability of falsely discovering an effect. *Power* of the test is the probability of correctly detecting an effect[17].

Hence, use of statistical technology needs to be done with competence and care. However, there is immense and growing demand for application of this technology, and – while extremely competent statisticians and mature theory and methods exist and continue to be developed – the majority of people and scientists using statistics are not *statisticians*[18]. Most *statistical science* consists of scientists using statistics, at varying degrees of sophistication and rigor, within their own disciplines. This may be even more true of statistics done beyond academia.

Regrettably, John Ioannidis (2005) concluded that, in <u>modern research, the majority of published findings are false</u>. In 2012, researchers from Amgen tried to reproduce the successes of 53 landmark cancer papers. They reproduced a mere 11% (Economist 2013, Begley and Ellis 2012, Baker 2016). Bayer HealthCare arrived at a 25% reproduction/validation (Prinz et al., 2011). In such egregious cases, we could well be better off without having any of the studies. Better to be uninformed than badly misinformed!

However, these error rates in excess of 75% – fifteen times the *conventional 0.05* – are far beyond the level inherent in statistical testing. What is going on here? Aside from a basic lack of rigor due to wide scale use[19], modern statistical methods and use have become more prone

---

[17] Power is a function, increasing with effect size as well as tolerance of type 1 errors, and reduced by stronger noise.
[18] Meaning those with degrees in statistics, those developing statistical theory and methods, and so on.
[19] Due to myriad defects in practice, *statistical science* is undergoing a full blown '*crisis of reproducibility*' (Nuzzo 2014, Leek and Peng 2015, Baker 2016). Although some blame the p-value for this crisis, analogous to a court trial (Wheatley and Sornette,

to false discovery, at times resembling *statistical witch-hunting* (Wheatley and Sornette, 2019). Some related factors for this include:

- Forgetting statisticians, it could well be that most statistical science is not even being done by humans: with the *big data* and *data science* fashions, algorithms scan high dimensional volumes of data, applying batteries of models and tests in a largely indiscriminate way.

- Practice of *multiple testing, overfitting, and data-mining/snooping* amplify false discoveries.

- *Positive-negative asymmetry*: Focus on success also affects academia. Indeed, 'we' accentuate and publish significant (positive) results without fully controlling for false positives—often requiring meta-analyses, and skewed by relevant negative results being unpublished and hence unknown.

Now that big data, artificial intelligence and machine learning are being hyped as resources and tools in big and growing scientific and commercial research, the fight against harmful 'lucky' false discoveries is a never-ending quest. This requires good statistics, as well as good old rigor and discipline across the full scientific process. We revisit this in discussion 6.4.

The investment industry provides a prime non-academic example. Indeed these issues muddy the waters, and allow the investment industry to thrive. For instance, investors choose to invest in funds based on their apparent skill: with *alpha* measuring excess return over a certain time period relative to its peers. Accounting for myriad statistical issues, in contrast to analyses by funds and investors, Barras et al. (2010) found that 75.4% were in fact *zero-alpha*, with stock-picking skills just sufficient to cover trading costs and expenses, 24% were found to be insufficiently skilled to cover the fees, and merely 0.6% were found to be skilled.

Further, automated *back-testing* and optimization of trading strategies on vast amounts of data, leads to an illusion of skill. This plagues quantitative trading firms, who can always find a strategy with superior performance on historical data, which then disappoints when deployed in real time (*out-of-sample*). A comical example is provided by D. Leinweber, a portfolio manager who "sifted through a United Nations CD-ROM and discovered that historically the single best prediction of the Standard & Poor's 500 stock index was butter production in Bangladesh." As a practical joke, he published this finding, and claims that people still earnestly contact him, trying to exploit this miraculous correlation (R. Sullivan, 1999).

## 4. How to play a game: an *evolutionary Greek operating system*

### 4.1 Framework

While the static skill-luck spectrum may be adequate for casino games, and even sports, when it comes to real world games we must consider a richer, dynamical and multi-dimensional process[20]. We thus dedicate ourselves to first developing the necessary systems view to be able to address real world problems. For this, we consider an instance of a *game*, visualized in figure 6, played by an *evolutionary learner*. The game has a structure, with luck (Greek 'sigma'), and skill (Greek 'mu' represented by M) characteristics. The player wishes to optimize their outcome, under risk, as well as uncertainty about the game itself. The adaptive learning consists of a reward-strategy feedback loop, and its refinement. We imbue the player with Greek rationality *logos*, to avoid misattribution of luck as skill. The perception of the strategy

---

2019), the lifecycle of a statistical study comprises various activities (Leek and Peng, 2015), with the decision of guilt being (one of the) last. Rigor is required throughout and false findings can arise from errors at any step. Apparently, few studies even rationalize the type 1 error level that they use – which should be case-specific, e.g., being informed by the degree of surprise or disruption associated with the test. This situation compelled the American Statistical Association (Wasserstein and Lazar 2016) to emphasize some modest minimal standards. There are many references on relevant good statistical practice (e.g. Zweig 2014, Lo and MacKinlay 1990, Sullivan et al. 1999, Bailey et al. 2015 and White 2000).

[20] Please note how the concept of the skill-versus-luck continuum only gives static information on an aggregated level. It does not take into consideration the fact that the game actually needs to be played by individual players, nor does it explain how this game may be played or how a strategy may be optimized.

and reward is, however, subject to cultural conventions/norms (*ethos*) or – most generally – paradigms, which are endogenous, and can shift. The performance and game itself may change with exogenous environmental conditions. It may occur that the potential to improve, based on exploitation of a given paradigm is exhausted. In this case, intensified exploratory learning can identify a new *local paradigm/optima* to refine and exploit. This forms the *mutation* part of the evolutionary learner.

4.2 Environmental conditions: stationary and non-stationary cases

*Environmental conditions* are defined as external: they can neither be affected nor predicted by the player but may impact the gameplay. When the environment is sufficiently stable, such that gameplay is stationary, the process can be described probabilistically and fully and directly optimized based on past performance (statistical inductive reasoning[21] works). When non-stationary, a deductive approach is needed, and the system is managed through the use of suitably realistic "toy" models, scenarios, and heuristics -- as well as the introduction of redundancy, and assessment of resilience. As a consequence, in this case, information from the past must be selectively used and adapted to future scenarios -- including novel ones.

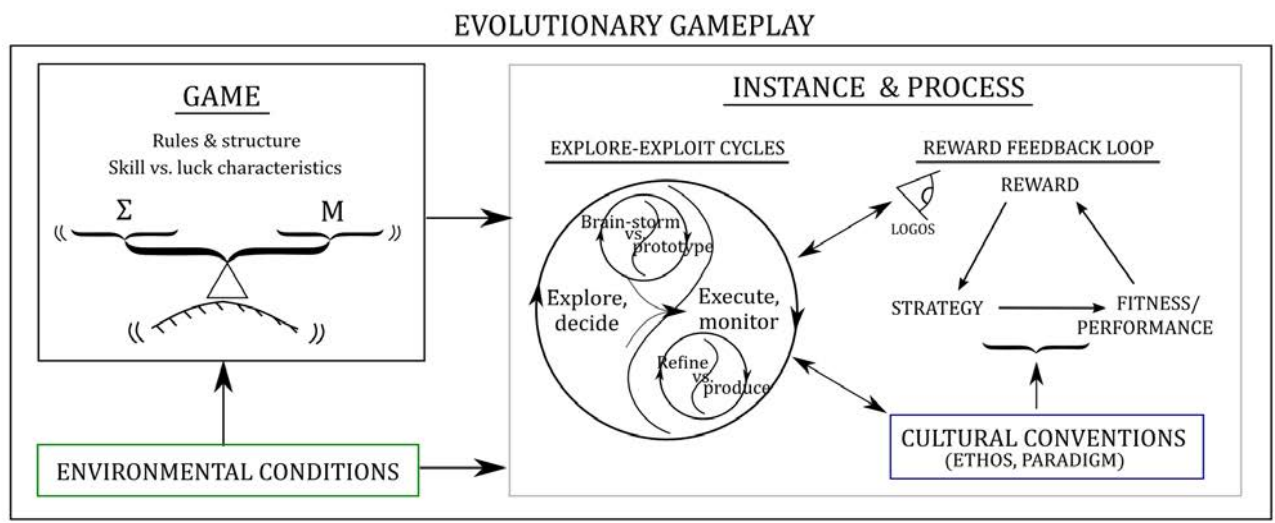

*Figure. 6: A game and an instance played by an evolutionary learner – described in 4.1 through 4.5.*

4.3 Cultural conventions & paradigms

Cultural conventions are defined as internal to gameplay and influence how fitness of a strategy and reward are perceived. E.g., conventions could affect the frequency used to sample fitness and adapt the strategy, whether the fitness is assessed at the individual or an aggregated level, how the role of luck is discounted in the award, if externalities are taken into consideration, etc. (Dorner, 1997). This is reasonable as characteristics such as risk and ambiguity preferences (e.g., view towards exploration of the unknown), delay of gratification, and collectivism vary by individual, across populations, and over time. Notably, the smaller the epistemic uncertainty (e.g., due to development of science), the more the player will aim to control and optimize. Some cultural paradigms may lead to a more fatalistic approach. Additional aspects and processes of this framework are further detailed in the following sections.

4.4 Navigating into the future as an *adaptive system*

Here, we want to incentivize skillful navigation into the future. Navigation needs an operating system. We imbue our evolutionary learner with five component processes: 'observe, decide, execute, challenge, and explore'. There is a large body of academic research on each of those topics. In the following, we will discuss the insights that are most relevant, based upon the work of Tetlock and Gardner (2015)[22] and Kahneman (2011), and references therein.

---

[21] General conclusions can be drawn from specific, individual, historical cases.

Observe: *Broadly decipher the role of skill, luck, and environmental changes*. In the non-stationary game, and with both aleatoric and epistemic uncertainties[23], sound observation of the relation between process and outcome is key. Consistency is important, which means that one has to neutralize irrelevant stimuli. This can be done by complementing as well as challenging a personal model of the world -- from which a subjective inside view condenses -- with objective outside information, or benchmarks. In this process, one has to challenge the internalization of observations with rules of thumb-- simple "back-of-the-envelope" calculations that enable a sanity check, or "smell test"[24].

Decide: *Strive for independent and diverse viewpoints during exploration of alternatives*. The key principle here is to profit from the *wisdom of the crowd* effect (Surowiecki, 2004) – whereby diverse independent viewpoints prove valuable in identifying solutions to complex problems[25]. It is essentially the law of large number and central limit theorem in action. In groups with interacting individuals, the basic wisdom of crowds cannot be assumed and needs to be remediated by dialectic and organizational mechanisms[26]. Indeed, group dynamics can strongly impact decision quality. Hence, during exploratory problem-solving, a consensus should not be the goal but rather a warning flag of detrimental groupthink. Indeed, during the Bay of Pigs fiasco, the Kennedy administration fell into this mode, losing independent critical thinking (Janis, 1972; Tuchman, 1984). Learning from this failure, dramatic changes were made, leading to better resolution of the Cuban missile crisis.[27]

Execute: *Unite behind a common goal, while allowing local initiative for adaptability*. Once a strategy is decided, the next step is execution. Napoleon, who won more battles than Hannibal, Alexander the Great and Caesar combined, mastered adapting plans extremely quickly: "So I have lived through general views, much more than definite plans. The mass of common interests, what I thought was the good of the very many, these were the anchors to which I remained moored, but around which I mostly floated at random."[28] Von Moltke, the famous 19th century Prussian military strategist, said, "*In combat, no plan survives contact with the enemy*". Mike Tyson came to a similar conclusion: "*Everybody has a plan, until they get hit*". The principle of *Auftragstaktik*, from German military history, is instructive: '*War cannot be conducted from the ivory tower*' -- tactical decisions should be made locally, allowing for rapid well-adapted decisions. But local tactics must be aligned with the global strategy, which must be regulated by clear goal-oriented orders from high-command, leaving initiative with local decision-makers. '*this is what needs to be done*', '*for the following reason*', '*I don't care how*'.

Challenge: *Constantly monitor, and ruthlessly scrutinize both successes and failures*. Any strategy executed in a dynamic environment must be constantly adapted, through *observation* and *decision* -- expectation is checked against observation, discrepancy is analyzed, leading to an adapted model (Seligman et al. 2013, 2016)[29]. *Challenging*, by conducting postmortems on failures and falsifying successes, makes the system adapt[30]. Like in execute, also here, a delicate balance between scrutiny (*exploration*) and continuation (*exploitation*) is needed. Sometimes, the best intervention is the act of deliberately doing nothing, following Hippocrates' "*Primum non nocere*"', first do no harm. A *light touch* approach can help avoid *analysis paralysis*. E.g., in contrast to arguably over-controlling invasions of Iraq and Libya by the US, the Ottoman and the British empires at their peaks knew to allow autonomy and to nudge towards balances of power between the many co-existing communities (Hellerstein, 2016).

---

[22] 'Navigating into the future' has been the research subject of Philip Tetlock et al. for more than 30 years (Tetlock and Gardner, 2015). In 1984, they started organizing forecasting tournaments to identify the distinctive features of expert judgements and predictions, and separate luck from skill in prediction – sometimes involving several thousand volunteer forecasters. These projects serve a dual goal: providing a real-life laboratory to study forecasting; and to use wisdom of the crowds for an optimized 'forecasting machine'.
[23] Aleatoric uncertainty is something you don't know but, at least in theory, is knowable -- like an engineer opening a mystery machine to find out its workings. Epistemic uncertainty is unknowable (relative to a given model), and in simplest form, means separating the signal from the noise.
[24] A simple often neglected technique is the use of independent base rates: E.g., studying an investment opportunity in a start-up, a venture capitalist should start from an independent assessment of the survival rate of comparable companies, and update this prior once specific inside information is available.
[25] Tetlock et al. argue that diversity of knowledge and opinion should be optimized. They call this applying 'the dragonfly view (Dragonflies have perhaps the most advanced color vision, due to large numbers of light-sensitive opsins ).
[26] Viewpoints should be carefully compared and dissected. And, one should avoid anchoring to the opinion of those who speak early and assertively (Kahneman, 2011).
[27] After the inquiry, skepticism was the new watchword. Experts were challenged and new advisors were brought in. JFK would often leave the room to let the group discuss more freely. This paid off with the successful handling of the Cuban missile crisis.
[28] Napoléon Bonaparte, Introduction to his biography "Les misères de Napoléon" by Lorenzi de Bradi, Editions Tallandier (1934).

With *observe, decide, execute and challenge*, we have the necessary ingredients for an intelligent adaptive machine. In the next part, we significantly enhance the intelligence by adding the key learning dimension: exploration.

4.5 *Exploration*: evolutionary learners are open to breakthrough mutations

As change (smooth, abrupt, and cyclical) tends to be the rule, any evolutionary system needs to continue re-optimizing to avoid stagnation and decay. And to avoid local optima, we need to explore, go beyond the trodden paths, and *play*. Stanley and Lehman (2015) explain that only by pursuing our curiosity without any objectives can true creativity be unleashed[31]. Perhaps they would agree that the pursuit of truth, knowledge, and prosperity is helpful.

More fundamentally, and already littered throughout the text, is the transcendent *exploration-exploitation paradox*[32], or rather *trade-off*. It is perhaps the most highly evolved ('trained') dimension of intelligent decision-making. E.g., consider that under hunter-gatherer subsistence, a regime occupying at least 90 percent of human history, it was perhaps the central problem: Indeed, prosperity and survival often depended upon balancing venturing into the unknown, with its potentials and risks, versus relying on familiar production methods. Not until the agricultural revolution, over 10'000 years ago, did we begin to learn to become political animals, within an organized/hierarchical system. In essence, it is equivalent to an evolutionary mutation-reproduction tradeoff[33]. Naturally, it forms the basis of learning in artificial intelligences (Kaelbling et al, 1996). It is a universal traditional dictum (e.g., enshrined centrally within Taoism, with the practice of balancing "qi"). Further, it is a problem that organisms, animals (Caraco, 1980), human individuals (Cohen et al, 2007), firms (innovate or stagnate; "managerial ambidexterity") (Wilson et al, 2014), strategic investors (invest in VC?), and societies alike need to navigate (Banks, 1994; Mehlhorn, 2015).

Collecting these ideas, we return to the game and evolutionary player in figure 7: The agent monitors (*observes*) the visible part of the system, attempting to discern luck from skill in a rational and objective way. Further, the dynamic environment alters the game, and potentially the balance between luck and skill (e.g., under highly variable winds, the lucky sailboat racer thrives). The agent needs to carefully infer the dynamics and contribution of luck, which are – at least partially -- beyond the field of vision. The system goes through explore (*decide*) – exploit (*execute*) cycles, alternating divergent open thinking and then unification to a common goal. A refinement loop embeds the same dynamic on a sub-scale (*monitor*), within the execution regime. Further, embedded cycles of disciplined convergence are needed within exploratory regimes: e.g., to select promising ideas and construct prototypes. Indeed, this is essentially the yin and yang of Chinese Taoism. This dynamic is illustrated in figure 7: In trajectory I(b) the system is tuned towards a local optimum and reaches a plateau. in I(a), a high peak was achieved but then decayed, e.g., due to excessive top-down control (poor *execution*), or environmental change. As the growth potential for the regime reaches its end, uncertainty proliferates, and – in this case -- intensified exploratory activity leads to a breakthrough. The mutation/shift from paradigm I to II takes place. Extinction was another possibility. In the new paradigm, operation is continually refined, and the system overtakes its peak performance from the previous paradigm – as visualized by the potential diagram, which is subject to future movement and uncertainty.

---

[29] Learning requires doing and doing requires execution. This is what Tetlock et al. (2015) call, 'mastering the error-balancing bicycle'. Andy Grove, the former CEO of Intel put it, 'constructive confrontation' should be practiced. This is echoed by Ray Dalio (2017): "In order to be successful, we have to have independent thinkers - so independent that they'll bet against the consensus".
[30] Tetlock et al. (2015) put it as follows: Don't try to justify or excuse your failures. Own them! Conduct unflinching postmortems: Where exactly did I go wrong? … Also don't forget to do postmortems on your successes too. Not all successes imply that your reasoning was right. You may have just lucked out by making offsetting errors. And if you keep confidently reasoning along the same lines, you are setting yourself up for a nasty surprise.
[31] They call this the 'Myth of the Objective': "objectives actually become obstacles towards more exciting achievements, like those involving discovery, creativity, invention, or innovation—or even achieving true happiness… the truest path to "blue sky" discovery or to fulfill boundless ambition, is to have no objective at all."
[32] Balancing costly, uncertain and risky learning with the safer conservation of "locally optimal" refinement and production.
[33] In biological evolution, for example, there is a progressive accumulation of small so-called "neutral" mutations that have no apparent effect on the fitness due to the huge redundancy of metabolic chemical reactions. Then, when a threshold is reached, the last minor mutation triggers a cascade of changes, which may lead to a new species (Wagner, 2015).

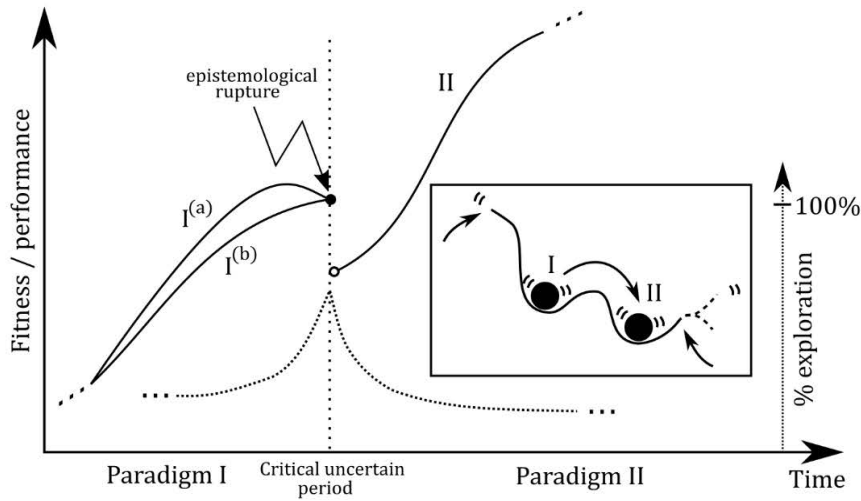

*Figure 7: Dynamic view of the evolving learner through 'paradigm change' and exploration cycles -- described within the text.*

### 5. Measuring & rewarding process performance

#### 5.1 The three cases:

Now that we have grasped the nature of evolutionary gameplay, providing a rich model for real world problems, we are in position to carve out three complementary ways of observing and hence rewarding merit. The specific properties and appropriate conditions for use of the measures are given. These three ways target on different measures of outcome:

Raw outcome ('waiting for the cream to rise'): This is the base case, where a process is evaluated based on a single snapshot, most typically at its stopping time. This is closely related to Dorner's ballistic action (Dorner et al. 1990). Where no measurement exists, adoption of outcome measurement would likely be an improvement. The simplicity of this approach – with e.g. success/failure objectively observable, not requiring assessment of process quality/merit, which is likely to be difficult, cost- and time-consuming, and prone to errors – makes it attractive. However, it is very noisy, relying on a single observation. In technical terms, a successful outcome might not say much about the actual quality of the underlying process – statistical power to significantly distinguish luck from skill is low. Inference may be skewed by changes of paradigm within the process. The following approaches are more sophisticated.

Risk-adjusted/aggregate ('separating the wheat from the chaff'): Here an instance of a process is evaluated over its path to see how the outcome was obtained. Under sufficiently stable conditions (environment), one can then use statistical experience to account for the role of luck leading to the final outcome. Hence this extends the raw outcome measure to be risk-adjusted. Further, if separate (ideally independent) instances of the process exist and can be observed, an aggregate measure can be used (e.g., % successful). Better discrimination between luck and skill are hence anticipated, relative to the 'sample size one' outcome-based approach.

Prospective: ('scouting future stars') This encompassing case addresses a changing environment, as well as changing internal paradigms – i.e., where history is not fully and directly representative of the future but, through deductive reasoning, can be assimilated to anticipate future scenarios. The process is hence evaluated by assessing the past, present and future, and its merit should -- in part -- be evaluated based on capacity to evolve. This allows for risk assessment going beyond scenarios and events that have occurred. Raw outcome and risk-adjusted measures may be used, where quantification allows. However measurement here is more about qualitative comparison of a process with the principles of an evolutionary system defined above in 3.

To reflect on this important *prospective* case more broadly, drawing inspiration from nature, Seligman et al. (2013, 2016) conjecture that prospection is the central and singular[34] organizing feature of human cognition and action. In view of this, humans should not be called *Homo Sapiens* ('Wise Man') but rather *Homo Prospectus* ('Forward Looking Man'), for whom consciousness is the generation of simulations about possible novel futures. Going further, Jeff Hawkins (2004) proposes a neuro-biological foundation, hypothesising that cognition is a feedback/recall loop to develop recognition and prediction in a bi-directional hierarchy: this "memory-prediction" theory claims that our brain is thus constantly trying to predict and then compare, predict, compare, learn, adjust, predict and so on. In related thinking, Karl Friston et al. (2006, 2012) posit that biological systems strive to minimize the differences between expectations and sensory perceptions (Perrinet et al., 2014).

To fix ideas, the three measures are plotted in a mind-map (figure 7), and related to the three types of operating systems, each with a higher degree of "intelligence" (Gell-Mann, 1994). The first one is direct adaptation: like a thermostat, which simply reacts when a certain threshold is crossed ('too cold, too hot, too cold'…). This most basic management system corresponds to our *outcome based* case. Next is a more intelligent *expert system*, which is a static internal model, like a decision tree, based on historical expert experience: 'if …, then …, else …'. The emphasis here is on control, and the belief that systems can be controlled with sufficient knowledge of historical input-output relationships. This corresponds to our *risk-adjusted* case. Finally, the most intelligent operating system is a *complex adaptive system*, which is an expert system that can learn from a changing environment by a process of mutation and selection, like in biological evolution: It has a set of built in rules (*scheme, gene, meme*). There are fitness criteria that allow for selection, and bad schemes become obsolete. Additionally, there is variation through reproduction and mutation (*exploitation-exploration*). Selection and variation are the basic ingredients that allow for the system to learn and adapt dynamically to its changing environment.

---

[34] We conjecture that other animals are also capable of conceptualizing exploration, extrapolating existing evidence for flexible planning (Kabadayi, C and M. Osvath, Ravens parallel great apes in flexible planning for tool-use and bartering, Science 357, 202-204, 2017).

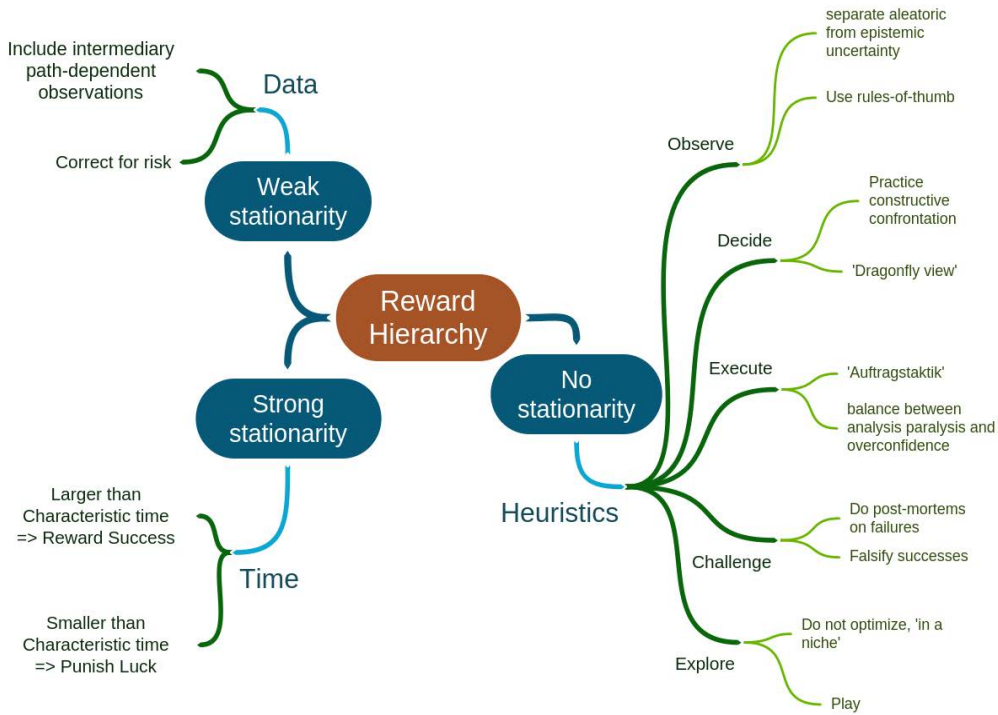

*Figure 8: The reward hierarchy: Three complementary branches representing fundamentally different approaches towards reward. In case of strong stationarity, (life-)time should be used, for weak stationarity, this should be data and when there is no stationarity, one should apply heuristics.*

### 5.2 Separating skill from luck: outcome and risk-adjusted measures

Short term success is often rewarded, irrespective and potentially at the detriment of long-term fitness. We strive for more meritocratic systems. Here the efficiency of the raw and risk-adjusted measures to discern quality is evaluated by simulation study.

### 5.2.1 Luck and skill model

We now move onto the problem of separating luck from skill under raw outcome observation. Consider a certain agent who uses a combination of skill and luck (exposure to which is affected by risk-taking) when completing a process (playing a game). Following Shockley (section 2), the distribution of the outcome of this task, at a certain point in time, is log-normal. In that case, its time dependent evolution can be represented as a stochastic process such as a Geometric Brownian Motion (GBM):

$$\frac{dS_t}{S_t} = \mu\, dt + \sigma\, dW_t \qquad (2)$$

where the percentage change $\frac{dS_t}{S_t}$ in outcome of success, in an infinitesimal time-step dt, is fully characterized by the percentage drift, $\mu$, which represents the skill part, and the percentage volatility $\sigma$, which here embodies the luck. The stochasticity is introduced by $dW_t$, which is the increment of a Wiener process (Brownian Motion).

Following a GBM, the excursion after a time T is typically:

$$\ln[S_t] \sim (\mu - \frac{\sigma^2}{2})T + x_i\, \sigma\, \sqrt{T} \qquad (3)$$

where $x_i$ is a random variable of zero mean and unit variance. The first term $(\mu - \frac{\sigma^2}{2})T$ in the r.h.s. of (3) is the cumulative effect of the skill component and the second term $x_i\, \sigma\, \sqrt{T}$ is the luck part. The term $(-\frac{\sigma^2}{2}T)$ in the skill component is the result of multiplicative noise in the stochastic process. For most applications, this correction term is small, so let it be zero.

There is a *characteristic time* T$^*$ at which skill and luck will contribute equally to the outcome of the GBM process. At that time T$^*$ satisfies $\mu T^* = \sigma \sqrt{T^*}$, yielding:

$$T^* = \left(\frac{\sigma}{\mu}\right)^2 \qquad (4)$$

For times T smaller than T$^*$, the luck component dominates, and the process is diffusive or random; when the outcome is evaluated at times larger than T$^*$, skill dominates, and the process is drifting. Hence, when evaluating a drift process unconditionally, based on a single observation, only time can divide between skill and luck. Only once beyond the characteristic time (4), the outcome of the process will be a good indicator for skill. At shorter times, luck dominates. More broadly, the characteristic time differentiates *Minyi* and *Minxin*, of the philosopher Mencius[35]: the first refers to policies based on fickle short-term thinking, while the second captures a long-term holistic view (Weiwei 2017). To consider some eminent views on the difficulty of estimating skill in a GBM type process: Merton, 1980 stated, *"… to attempt to estimate the expected return on the market is to embark on a fool's errand"*, and Markowitz and von Dijk, 2006, *"We know no procedure to put data in and get "correct" expected return estimates out."* Ambarish and Seigel, 1996, show that it can take years to estimate expected returns for a stationary process to a specified confidence level. Chopra and Ziemba, 1993, discuss errors in means on portfolio choice and show their relative importance.

5.2.2 Time as a divider: *waiting for the cream to rise*

For the raw outcome observation case, we assess the performance of the characteristic time heuristic in separating skill from luck: We consider a heterogeneous population of N agents, performing some process, following the GBM (2). Each agent has skill $\mu_i$ and luck $\sigma_i$ (e.g., due to different risk-taking levels). The $\mu_i$ and $\sigma_i$ are distributed according to two lognormal distributions LN($\mu_{skill}$, $\sigma_{skill}$) and LN($\mu_{luck}$, $\sigma_{luck}$) [36].

After a time-step $\Delta T$, the agent's realized success can be calculated from equation (3): with skill part $(\mu_i - \frac{\sigma_i^2}{2})\Delta T$, and luck proportional to $\sigma_i \sqrt{\Delta T}$. We then split the population into outcome deciles (each containing N/10 agents) according to their realized outcome after time-step $\Delta T$. We then calculate the average of the true skill $\mu_i$ and the luck $\sigma_i$ for each observed decile. The simulation is repeated for different $\Delta T$, which we will call the "vetting period": the time needed to assess the qualities of, or to "vet", the agent. The results for four different populations, with increasing heterogeneity, are presented in appendix A.1.

From here, and referring to figure A.1, considering the *reward problem,* two solutions appear:

1) '*select on success*': rewarding the top deciles, hopefully containing skillful agents, or

2) '*de-select luck*': rewarding the middle deciles and hence avoiding/'punishing' the most lucky – and perhaps those taking irresponsible risks. Indeed the top deciles contain the most lucky (and some skillful) individuals. Rewarding them is the classical raw outcome solution. The merit of this strategy, however, is strongly dependent on population heterogeneity and the vetting period with respect to the characteristic time.

Interestingly, the de-selecting luck approach depends weakly on population heterogeneity and vetting time. This can be clearly seen in the four middle subplots in figure A.1. By rewarding the middle deciles, we end up rewarding agents with a much lower relative contribution of luck, in fact explicitly ignoring the most successful and unsuccessful. Pluchino et al. (2018), based on an agent based model, also find that *the most successful agents are seldom most talented.*

---

[35] An eminent Confucian philosopher who lived from 372 until 289 BCE.
[36] To be clear, $\mu_{skill}$ is the mean and $\sigma_{skill}$ the standard deviation of the distribution of the skills of the agents, while $\mu_{luck}$ is the mean and $\sigma_{luck}$ the standard deviation of the distribution of the 'luck' of the N agents (allowing for e.g., variable risk-taking). The term $\mu_{luck}$ should not be confused with a drift term. It expresses the mean volatility. The same holds for $\sigma_{skill}$ which should not be confused with a volatility, but population variability.

Which solution is best? Obviously, given a long enough vetting period, one can just wait and take the cream from the top deciles. However, how does one know if one has waited long enough such that the observed success is not mostly by chance? And, what to do in practical cases of heterogeneous populations and short vetting period? For instance, for the problem of selecting financial investments, given that typically $\mu \sim 5-10\%$ and $\sigma \sim 20-30\%$ per year, the characteristic time T* given by expression (4) is between 4 and 36 years. Here the (unconditional) select-on-success approach is weak for holding periods less than 4 years, which is already white long. This illustrates the well-known fact that most superior performing financial funds enjoy significant luck, while the truly skilled ones are rare (Barras et al., 2010).

We now test performance of the above through the lens of a risk-adjusted measure: skill divided by luck ($\mu/\sigma$) called *Sharpe ratio* in finance. Results are shown in figure 8. It is striking that, for shorter vetting periods -- here less than one year -- the Sharpe ratio of the optimal allocation is almost flat. Then there is a shift from an allocation to the median deciles to an allocation to the first decile, forming a step-like optimal allocation function.

The lessons from our simulations are clear: Only if one has the luxury of very long vetting periods, can one separate skill from success in this way – in a stochastic and heterogeneous world, the cream rises slowly. For most real human activities, the broad distributions of skill and luck, together with the urgency of decisions, requires efforts to avoid the *Illusion of Success* that tends to blind and misguide us in the pursuit of sustainable successes.

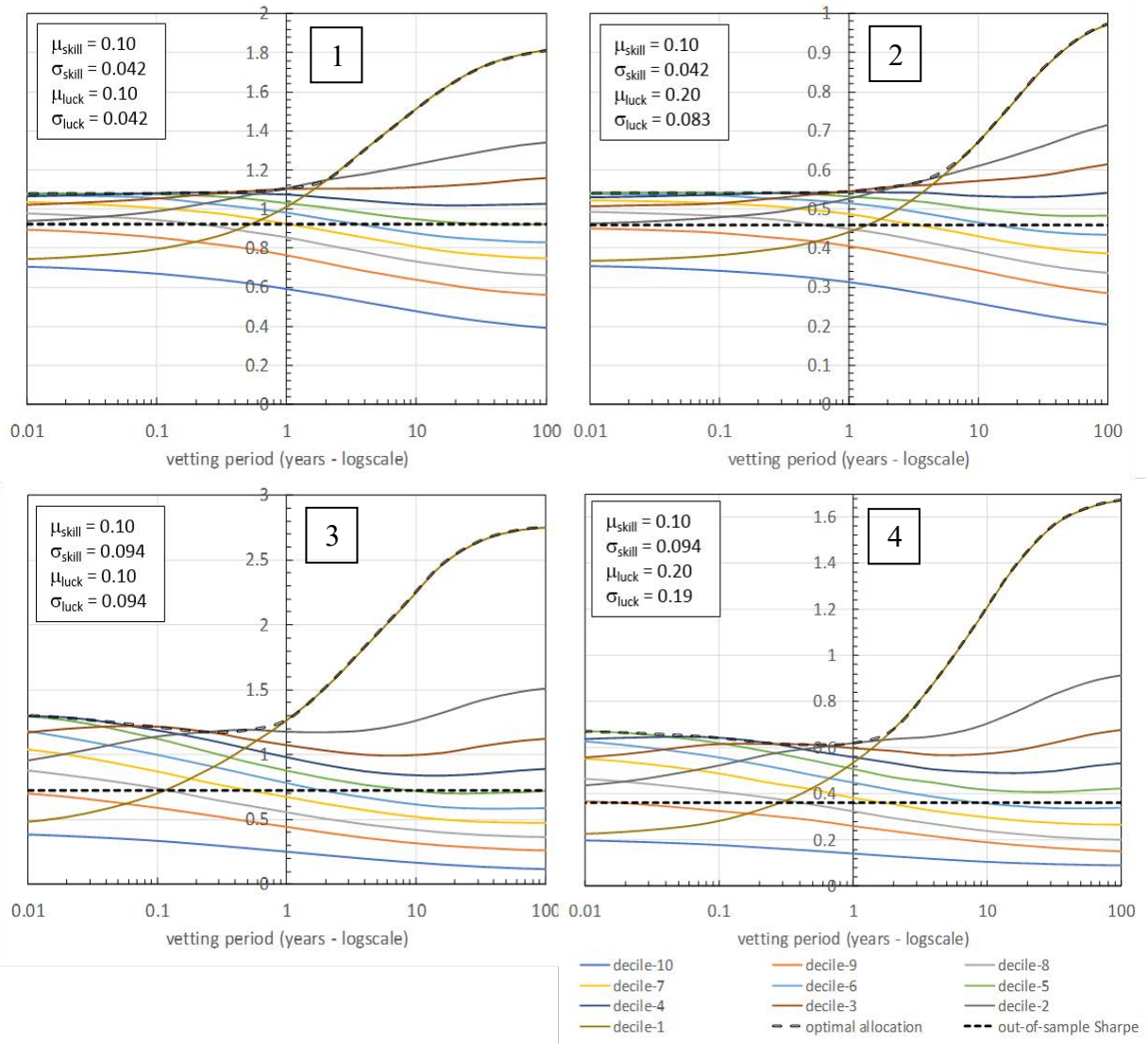

*Figure 9: Sharpe ratio (y-axis) as a function of the vetting period (x-axis, log-scale) for the different deciles. Each subplot represents a different population whose distribution of individual skills and lucks are log-normally distributed with parameters given in the insets of the panels (see Appendix A.1). The results for the different deciles are represented by curves with different colors. The horizontal black dotted line gives the Sharpe ratio for a one year out-of-sample period for the population and the black barred line gives the optimal allocation, which is defined as the winning decile for each vetting period. In all cases, the transition where select-on-success overtakes deselect-on-luck, is near the characteristic time, and largely in agreement with $(\mu_{luck}/\mu_{skill})^2$: being one, four, one, and four years for the numbered plots. E.g., for population 1, (upper left), deselect-on-luck is dominant for vetting periods up to one year, as deciles 3-5 are optimal. Beyond that, select-on-success wins.*

### 5.2.3 Risk-adjusted measures: *separating the wheat from the chaff*

We now consider the richer setting where 1) we are able to include intermediate results in our evaluation process, or even 2) consider multiple parallel instances. Considering the first case, and recapping: if the underlying processes are quasi-stable, the past is a good indicator for the future and a risk-adjusted measure is sound. To study this, we use the same simulation as above, but sample/record intermediate results over the period $\Delta T$ of each instance. This allows estimation of the drift (skill) and volatility (risk taken or inherent in their process) of each agent.

Many risk-adjusted measures can be considered, with the strength of volatility being an important factor. For simplicity, and appropriate for our model (eq. 2), we use the Sharpe ratio.

For our simulation study, this ratio is used to rank the agents and sort them in deciles, each containing N/10 individuals. For each of the deciles, we calculate the average true skill and luck using the parameter values given to the agents in the simulation. We use a vetting period of one year, and we gradually increase the number of intermediate observations from 2, 4, 8 … to 256.

The results (see figure A.2) are quite clear, even for a very crude statistic, e.g., using only two observations per agent path: The Sharpe ratio selects higher skill and lower luck agents, monotonously over the 10 deciles. Hence, by adding even a single intermediate measurement, in addition to the raw outcome -- which allows for a very crude estimate of the volatility based on two observations -- the typical smile-shape for luck, seen in the outcome-based study, completely disappears. This is very promising, although in practice one must take care so that the measurements are reliable and meaningful. If parallel instances are somehow available, then aggregate measures will "separate the wheat from the chaff" with even greater efficiency.

5.3 On reward itself: Dividing the dollar

By now, we have considered separating luck from skill and broader merit in general dynamical systems/games. We briefly consider the problem of distributing reward itself. In the first two cases, raw outcome and risk-adjusted outcome can be directly used for reward/compensation – subject to their relative strengths, as already covered. In the third case of *prospective* reward the problem takes on a new forward-looking element. This question arises when hiring, when scouting prospects for professional sports, when choosing a mate, and so on.

However, thus far, we have swept another issue under the rug, by considering a player, agent, or operating system to be rewarded. However, the fact is that, in many cases, the relevant activities involve groups of people, who through individual contribution as well as interaction determine the results. But identifying the true contribution of each member to the group is devilishly difficult, especially when members have different complementary domains of expertise, and in the presence of uncertainty (and unpredictability in the prospective reward case). Who has not been part of a project, where the sum of the self-attributed contributions over all members of the team was much above 100%? Even stringent quantitative criteria may be difficult to align with personal subjective perceptions and, often, do not apply in many situations with "soft" qualitative dimensions.

In fact, the way the recompense is allocated also constitutes a crucial component to success, while it leads to conflicts and to disincentives if badly implemented. Prospective solutions require both qualitative and quantitative assessments that are difficult to fit into objective top-down directed criteria. The mechanism should support successful self-organization, growth, and merit.

In "Dividing the dollar", de Clippel et al. (2008), followed by Tideman and Plassman (2008), provide a general and powerful methodology to divide profits in a partnership of three or more individuals in the absence of rigorous quantitative criteria. The principles behind the distribution mechanism are the following. First it should be *impartial* (no partner is able to affect their own share), second it should be *objective* (a partner only gives input about the others' performance), and *consensual* (the division of shares should be consistent with input of all partners). For three partners, there is a unique solution, and for more, a family of *optimal* division rules. Operationally, this methodology is very practical, intuitive, and transparent.

Such a framework can incentivise team members to prioritize efforts and skill, rather than counting on signalling, bargaining and strategies to get rewards. And clearly, outcome, risk-adjusted, and prospective assessments can form a valuable basis for the reward process.

## 6. Discussion and applications

Here we apply the ideas and means developed to diverse important fields to address the *fair reward problem*. Recalling the static representation in figure 3, hard work and clear thinking is needed to see where the actual balance between skill and luck lies. Further, dynamics, environmental and cultural factors are relevant. A system's view such as ours (figure 6 and 8) becomes necessary as prospective evaluation of these systems is important.

### 6.1 Content aggregators: the *Amy & Frank* algorithm

As an introduction, consider the story told in the "Hang the DJ" episode of visionary series "Black Mirror"[37] (spoiler warning!): Characters Amy and Frank who, so far unsuccessful in their romantic lives, signed up for an advanced speed-dating system. The game starts, and they meet a number of partners, with mixed results. Eventually they are paired together and strongly connect. However, the system then tragically obligates them to move onto new partners. Across many rounds of the game, they fail to find one-another. Eventually, they succeed, and – in a dramatic breach of the strict rules of the game -- run off together, breaking free of the system. The frame zooms out to a summary revealing their reality to have been one of 1'000 independent simulations. In 998 of the cases, they obtained this romantic escape outcome – indicating strong compatibility. The frame zooms out again revealing that the entire test had been conducted on a smartphone, at a bar, where (supposedly real) Amy confidently approaches Frank.

This story is useful for delineating the three methods of section 5: Each simulation returns an *outcome*. They then provide an *aggregate outcome measure*, which can be thought of as accomplishing the goal of the *risk-adjusted measure* – by averaging across simulations, they consistently estimate quality of match. A risk-adjusted measure here would involve tracking the volatility of some quantities within the development of a given instance. Finally, the simulation is prospective: the model-versions of Amy and Frank are tested in a number of novel scenarios.

If only we could run such a realistic Monte-Carlo algorithm to solve the fair reward problem in other contexts; then the fair reward could be readily obtained by repeated simulation, and all counterfactuals could be known. However, in fact, with current technology, some important cases exist where this can be attained. In particular, consider any system that aggregates content items – whether they be ideas, tweets, news, videos, etc. – in a social, organizational, academic, or political context. These are major domains for society, commerce, knowledge, governance, and so on. However, as established in 2.2, they are plagued by viral winner-take-all phenomena. The Amy and Frank algorithm provides a solution. As the academic prediction market is discussed here separately, we exemplify by considering the major news aggregator, Reddit:

On Reddit, users submit content – typically with a link to an external article – and can then vote these items up or down. The top-ranked items are then seen by many users, receiving more attention, and being circulated elsewhere. Unsurprisingly, searching for a certain keyword, one will find several similar posts, "waiting to be discovered", with most having minimal success, but perhaps a few of them having become viral stars. Amy and Frank's system is clear: For the appropriate vetting period, the count/ranking of items (tweets, comments, etc.) will be done separately in a number of compartments, with users/readers randomly assigned to each compartment. Items will have varying success, and the rankings will vary. After the vetting period, an aggregate measure of the item can be made, and rankings updated across compartments to reflect that.

To reiterate, such ranking systems are of fundamental importance: YouTube, Twitter, and Reddit are clearly within the top 10 (non-Chinese) websites. The ideas apply to search engines as well. There are many structural issues with current information systems and media (e.g.,

---

[37] Series 4, Episode 4, written by Charlie Brooker, directed by Tim Van Patten. (29 December 2017)

leading to "fake news", and polarization). However, a perhaps underappreciated mega-factor is the increasingly wild viral dynamics of ideas, which favor sensationalism and low quality content, and confuse and bewilder the population. This algorithm shows that we can recover some *wisdom of crowds* -- significantly reducing noise and increasing meritocracy – while still supporting a more sane and truthful aggregation of information. More meritocratic and moderate distributions of success would tend to result, inter alia supporting more content creators instead of just a few stars.

6.2 Accident precursors in critical industries: a risk-adjusted approach

In striving to assess merit beyond raw outcome, we consider important *critical industries* – electricity, nuclear power, aviation, etc. These objects generate tremendous value but can also be harmful. Indeed, quantification of risk of such systems is key for decisions (e.g., selecting energy sources), as well as liability in the case of negligence.

Challenges – of external, technical, and human origin -- to system safety take place, and systems and processes mitigate progression to more severe states. In many cases – e.g., in a nuclear power plant – there is a threshold effect whereby, if the event does not progress too far, consequences may be negligible, and the event may be considered an *incident* or even simply as a *vulnerability/degradation*. However, in other cases, the chain could have come close to being consequential (a *near-miss*); and in other cases, a full accident occurs. Hence, representing an event by a 0-1 *raw outcome* is reductive, and where accidents are rare, absence of accidents is a poor (noisy) performance measure. These partially realized accident sequences are called accident precursors, from which performance information can be gleaned (Kunreuther, et al, 2004). This relates to a risk-adjusted approach.

As an illustration, Three Mile Island, 1979, can be thought of as a major accident -- notwithstanding its quasi-absent release of radioactive substances -- because there was a significant chance of a large release (Sornette et al., 2018). More recently, was the case of Fukushima Daini which, when hit by the 2011 Tōhoku seaquake and tsunami, happened to retain an external power line, and hence did not fall into complete blackout, unlike the less fortunate Daichi site. Was Daini "better", or, to some extent, more fortunate?

Unlike most other sectors, precursor analysis within nuclear power profits from mature and detailed plant-specific probabilistic safety analysis (PSA) models, allowing computation of the probability of realizing a core damage accident, given some hazardous condition. As an aside, this approach, as well as overall regulation of nuclear power stations which is done on the basis of estimated frequency of core damage accidents, could benefit from more systematic consideration of severity of accident outcomes, yielding more complete risk information, which varies by site, design, and so on.[38] This approach allows counterfactual analysis, where the higher the conditional probability of accident, the closer the near-miss. Despite few bona-fide nuclear accidents, there are hundreds of non-negligible precursors (Sornette et al., 2018). If such severe precursors – where accidents were avoided by luck -- are not penalized, a culture of over-confidence and mis-attribution leads to the degradation of governance and of risk management. This process explains the maturation mechanisms underlying many catastrophes (Chernov and Sornette, 2016). These considerations apply to many other industrial sectors. The tragic Lion Air Flight 610 (October 2018) and Ethiopian Airlines Flight 302 (March 2019) accidents of Boeing 737 MAX aircraft provide instructive examples: In total, these accidents led to 346 dead, and will be very costly. Does the liability lie with Boeing, the pilots, or is it just bad luck? In the October accident, it was reported that the same aircraft experienced a similar malfunction the day before, and fortunately was corrected by an extra pilot in the jumpseat (Levin & Suhartono, 2019). Evidently this precursor was not analyzed in time to raise the alarm. John Casani, former Assistant Lab Director at NASA's Jet Propulsion Laboratory, proposed that a system like a prediction market, including incentives for whistleblowing / reporting of risk information, could have helped avoid mishaps experienced in their major missions due to knowable errors[39]. It has been claimed, that under tight competition, Boeing took a design decision to increase engine size and change their location, without fuselage re-design. This changed the aerodynamics, requiring an automatic sensor driven system to counterbalance (Leopold, 2019). When spuriously actuating, the system would drive the plane down, requiring hitting a circuit breaker to override and avoid a crash. Consistent with the two accidents, pilots were unaware of this (Gates, 2019). This is particularly shameful, where flight simulators (prospective approach) have long existed to train and test pilots.

---

[38] The INES rating does this partially, but in a rather crude and not fully consistent way, and is more intended as a communication tool. See analysis and discussion in (Sornette et al., 2018).

<u>6.3 Political evolutionary system: battling giants (equalizer & ossifier[40])</u>

It is instructive to consider *the state* but as one paradigm within an overarching evolving operating system, tasked with the *objective* to maximize prosperity, playing in a global power game. This encompassing object *requires as well as confirms* the full richness of our evolutionary framework, with its internal dynamics and culture, as well as external effects. It also fundamentally links relevant objects: society, taxation, and meritocracy. Elements of this *systems politics* view are discussed below.

The modern state *paradigm* can be defined by having evolved taxation, citizenship, and civilian armies, allowing for economies of scale as well as formidable military forces (Cederman et al., 2011). This has grown the state skeleton. Smaller and less structured societies were simply dominated, despite having potentially nice internal characteristics (this is part of the *game* and its *environment*). Internal conflict poses a threat as well. Sometimes to increase output, but also to buy internal stability, taxation is used to transfer from the 'haves' to the 'have-nots'. In essence, welfare/tax schemes need to be balanced with negative incentives and constraints (Mankiw et al., 2009). Inter alia, under increased tax revenue the state grows, imposing its structure, while attempting to maintain internal harmony – in the global game, it is not only about meritocracy.

Performance information (*observe*) is likely to support sound governance. However, one wants to optimize the performance of the system, not its measurability. For instance, the vetting period necessary for assessing the performance of a government may be far longer than what is desirable, say for a democratic system. Further, relevant information is often polarized, with nuance and even basic truth falling casualty. Section 6.1 provides a partial solution. Also note that this is being given as a prime example of where *prospective* assessment of evolutionary characteristics are needed.

Performance of the top level objective can often be questioned as, even in representative democracies, the decision-makers often have their own objectives and practice nepotism[41]. In part, this can relate to the weak link between awarded political power and merit: Democratic processes have been shown to be highly stochastic and often biased (Galam, 2012), and in any "winner takes all" system, the reward is clearly poorly distributed[42]. It is instructive to consider the *direct democracy* of Switzerland which is a shining example, and where 'they' are working on new designs to align stakeholder interests and limit corruption of power[43].

Currently, external and internal mega-shifts in technology, the economy, and society have almost necessarily rendered existing governance models sub-optimal. Internal rising inequality will put great stress onto society, which will become less tolerant of nepotistic government rent-holders. External power struggles seem to intensify. To touch on the key *exploration/decide-exploitation/execute* dynamic, which becomes so key around crises: Successful government is typically – within a given paradigm – a late adopter of various innovations, as exploration can be risky. However, in the event of political crises, survival of a polity depends on the ability to intensify *mutation/exploration* and transform itself – and luck. The polity will need to effectively *decide* – avoiding politically-correct groupthink – and then unite behind a correct common goal and *execute*. Democracies will not be spared from this evolutionary challenge. Recalling *Agamemnon*, the bigger and greater the skeletal order, the more internal entropic forces work against it; corruption and politically correct orthodoxy are prime historical examples. This is what Mancur Olson calls 'institutional sclerosis', which is mainly caused by special interest groups and free-riders.

---

[39] Plenary speech at PSAM 14 conference at UCLA, 20-09-2018.

[40] By ossifier, we refer to the creation of entranched stagnating structures with the government, which wants to get big to dominate other giants in the global game. At the same time, it needs to maintain internal order (there equality is a big issue, as well as repression of conflicting groups within), which becomes increasingly difficult as it grows and ages.

[41] Responses to the 2008 financial crisis in the US and the 2010 sovereign debt crisis in Europe, in the form of bank bailout programs, quantitative easing and low interest policies largely benefited to wealthy equity and real-estate holders.

[42] Bush beat Gore, in the 2000 US presidential election, due to Florida, by a margin of just 0.009%, or 537 votes!

[43] Flexible majority rules, incentive contracts (dual democracy), two-stage unanimity rules, rotating agenda setting and agenda repetition in combination with flexible majority rules, minority voting, balanced voting, assessment voting and so on. In (Wiesner et al., 2019), democratic stability is interpred from a 'complexity science approach'.

We conclude by taking a narrower view on income and taxation: Corresponding to the edge-cases of section 2, it has been argued (Mirrlees and Diamond, 1971) that, in a pure meritocracy, where the social planner only observes income (raw outcome), a regressive tax best stimulates production and welfare. However, even if merit were directly observable, this ignores both luck and social stability. Indeed, income is often affected by luck, especially at extreme wealth levels. In an artificial 'lottery game economy' limit, with everything determined by luck, a large tax rate would not disincentivize 'work', and has the lucky subsidize the unlucky (Varian, 2001). Progressive taxation becomes the tool of last resort to recover meritocracy in a luck-dominated environment, reminiscent of our rewarding of middle outcome deciles in 5.2.2. In the bigger game, meritocracy can also be traded-off in favour of social stability.

This leads to an interesting but hard to implement idea: The role of luck varies by activity (figure 3), and risk-taking as well as externalities are in some cases subsidized. Hence differentiated tax rates are justified – exactly as insurance policy, where individuals are rated according to their risk. Effectively, "follow your dreams" is seldom a socially responsible approach, where risk sharing exists. In such a design, soccer super-stars would be taxed more than basketball super-stars[44], and financial professionals at a higher rate than engineers with similar skills[45].

6.4 Science and innovation: the explorer goes to the market

Indeed science (in a broad sense, including academic, state, and firm R&D) is where we look for innovative solutions to problems facing society[46], but also to conserve and refine the existing body of knowledge. Its dynamic balance lies at the heart of our evolving system (figure 6 and 8). But is the current scientific system well designed and adapted? For instance, the heterodox academy[47] is questioning if the predominance of progressive personality (and politics) in the academy may be problematic, e.g., leading to 'throwing out the baby with the bathwater' and lack of open-minded problem solving (*decide*). The role of luck as skill in statistical science was covered in section 3.6. Here, some features of the current system, and a new model are discussed.

The winner-takes-all mechanism plays an important role in science, *first mover advantages* create viral superstars, with large rewards given to few (Nobel prizes, large research grants …) (Simkin and Roychowdhury, 2018). However, the achievements are often a large team's work and benefit from a supportive social context as well as technological and conceptual maturations (Merton, 1961). This is discussed further in the next section. Not unlike media sensationalism, there is even a culture of social hypes where "well-dressed trivia" obtain the majority of the attention in top journals (Buchanan, 2009). Not only is this unfair in terms of rewarding scientific merit, but – more importantly – leads to an inefficiency in the assimilation and elevation of knowledge in the scientific body. A better indicator of merit of a scientist and work would control for clout/network-effects, and risk-adjust to avoid being skewed by publications that '*went viral*' largely by chance or by resonance with a lowest common denominator.

---

[44] A referee pointed out as a paradox: Roger Federer, was earning more than Novak Djokovic also when the latter was performing better. However, this can be explained by cumulative success over several years. Further, Federer's value to sponsors does not only depend upon his winning, but his perceived 'greatness', and effectiveness in advertisement..
[45] The Finance Wage Premium: the gap in compensation between the financial industry and the real economy, which cannot be explained by merit (Böhm et al., 2018) leads to a brain drain into the financial sector. The financial sector benefits from rent, associated with their specific position in the flow of money, as well as the focus on short-term success rather than true skill; and socialized loses.
[46] Richard Smalley, a Nobel prize laureate in Chemistry, famously developed a list of humanity's top ten grand challenges in the 21st Century: (1) energy; (2) water (more than 1B people lack access to safe water); (3) food; (4) environment; (5) poverty (3B people living on less than $2/day); (6) terrorism and war; (7) disease; (8) education; (9) democracy and (10) population (Smalley, 2005). The UN Secretary-General's High-level Panel Report on Threats, Challenges and Change, is rather different (https://www.un.org/ruleoflaw/blog/document/the-secretary-generals-high-level-panel-report-on-threats-challenges-and-change-a-more-secure-world-our-shared-responsibility.)
[47] https://heterodoxacademy.org/

It is a sociological fact that innovative discoveries are almost never published in prominent scientific journals and only recognized after much efforts from their authors: the system resists disruptive revolutions, preferring to accept incremental successes (Kuhn, 1970; Dyson, 1988) – falling into "*technical solutionism*" (Morozov 2014, Spiekermann 2017) and a brittle orthodoxy. According to Ohid Yaqub (2018), the pursuit of efficiency and a mechanical reduction of errors in scientific research may suppress the error-borne mechanism of serendipity. Based on the Merton archives, he collected hundreds of examples of serendipity in research: discoveries based on '*happy accidents*'[48]. Indeed, scientists should take risks and explore the unknown as needed. Further, the scientific body should be open to new paradigms, and aggressively reconcile new findings with the current understanding. This is clearly important for society. Rigorous fundamental research must also continue, with a perhaps surprising externality that many revolutionary findings came from detailed and incremental research where nature inspires a new technology (Sornette and Zajdenweber, 1999).

Considering the system: typically, research funding decisions are based on proposals that are reviewed by small and closed committees. The process has a *prospective* element in that the proposal is forward looking. In practice, it is rather inefficient and evidently favors incremental research, further entrenching the paradigm. Indeed, well established reviewers may have their personal activities disrupted by innovative works, without exposure to the upside. To improve, *market* mechanisms have been proposed. Let us develop some of the details.

A market for scientific research could be established, with scientists having funding to invest in proposals and ongoing projects of others (Bollen, 2018), with shares traded on an exchange. Proposals would be released like an initial public offering, with the writers of the proposal receiving a fraction, and the rest sold to the relevant community. Shares would then be subsequently traded. Participants would buy and sell shares in projects based on perceived merit. To enrich the exchange, financial options have been proposed (see Weinstein in Buchanan, 2009)[49]. Importantly, the market should include both forward science as well as falsification, which could be incentivized through short-selling. Further, regulators could monitor for unsafe or fraudulent activity.

To reflect on the profound potentials of this novel model: This design fills two needs with one deed, rewarding merit on both the proposal/production and review sides: high quality science and review would be reflected in portfolio value increase (money flowing to merit), leading e.g., to more real funding. The efficiency of financial markets would help the cream rise to the top[50]. Importantly, reviewers could then profit from the upside of exploratory proposals, rather than just suffering from the resultant disruption as a negative externality. Both the science and its peer-review would have motivation to be rigorous, transparent, and objective[51]**:** full disclosure

---

[48] when something has inadvertently been dropped, spilled, heated, exploded, forgotten about in pockets or drawers or laid to rest over holidays, contaminated or subjected to methodological blunders and/or equipment malfunction.

[49] Supporters of gradual increase in science would be option sellers ('short volatility, long theta'). As long as scientific progress is smooth and incremental, as would be their payoff (like a fixed income investment). The buy side of the options would be bleeding cash periodically in small amounts, but would be payed handsomely in event of a breakthrough.

[50] Here, the term 'best' refers to a ranking anointed by the choices of the scientific community. This mechanism does not necessarily avoid the emergence of social hypes and 'well-dressed trivia' that could obtain the majority of the attention. But see below the discussion on 'social bubbles'.

[51] A prototype of this design, called xYotta, has been developed at the chair of Entrepreneurial Risks in ETH Zurich, to provide a radically innovative way of how courses are taught, and group work is organized and evaluated. xYotta utilizes the wisdom of the crowd (the students in a University set-up) and market mechanisms to identify the best projects, ideas and talents. It has

of data, results, etc. would attract investors (funding). Investors would perform rigorous review, together with short positions, encouraging a productive dialectic of positive claims and falsification efforts. As a positive externality, the system would generate measurable information to construct aggregate and risk-adjusted success measures for both reviewers and works: making the ultimate decision on publish or not easier and more merit based. Through portfolio diversification, an intelligent balance between exploratory and incremental work would occur. This could relax publish or perish pressures, e.g., where a researcher could hedge risk of their own exploratory work by investing in other secure incremental projects.

Of course, potential market manipulation and regulatory issues require careful thought. A valid concern is that market bubbles may occur, leading to a concentration of funding. However – as embedded in our evolutionary system (figs. 5,6) -- such excesses, the "social bubbles", may enable great breakthroughs neither attainable nor with potentials foreseen through standard cost-benefit analysis (Gisler and Sornette, 2009; 2010; Gisler et al., 2011).

We conclude with some points on *observe* and *challenge*:

- Observe: Avoid being tricked by scientific (statistical or otherwise) malpractice, and the 'halo effect'/preferential attachment to well-known scientists. Don't shy away from 'prospective' deductive reasoning. Just because there is no data does not mean that there should be no research. "Not everything that counts can be counted".
- Challenge: Fund and publish replication studies and negative results. An outlier may be a 'paradigm-killer'. Automatically accept peer-reviewed falsification of a paper in the same journal [52].

### 6.5 Innovation: private sector & state

Here the *reward problem* for innovation between the state and private sector is briefly considered. On the one hand, former president Obama's "*you didn't build that*" offended entrepreneurs. But, egregious patents and state-supported monopolies exist, offering undeserved rent. Technological innovation relies on an existing science and technology base, which is largely a patrimony of humanity[53]. At the same time, advancing technology and knowledge is difficult and beneficial. Fair reward must be offered.

It is instructive to recognize the key role that a state can play in innovation (Mazzucato, 2013): forming part of a complex game along with financial capitalists and entrepreneurs (Janeway, 2012). In the United States, major innovations without state involvement even seem somewhat anomalous, where state means and purposes generated innovations such as: the modern Internet, semiconductors, nuclear energy, human genome research, liquid-crystal displays (LCD), global positioning systems (GPS), multi-touch screens, artificial intelligence with a voice-user interface, etc. Silicon valley (Blank, 2017) and even Google (Schachtman, 2010; Szoldra, 2016; Nesbit, 2017) were largely 'seeded' by state funds.

Given existence of incremental regimes, along with exploratory paradigm-shifts (figure 7), basic principles for fair reward can be drawn: In the realm of incremental developments, the relation between merit and reward is clear, and returns on such work should not be heavily taxed. However, as/where most exploratory work is subsidized by the public, reasonable taxation of its commercial exploitation is justified.

### 6.6: Financial risk management

Risk in financial institutions is largely assessed and regulated on the basis of summary quantitative risk metrics, determined by *internal models* (developed by and within the institutions). The systemic importance of the performance of these functions, and painful and

---

been rigorously tested by more than a thousand ETH graduate students in several courses and research projects over several years (Sanadgol and Sornette, 2017; Sornette et al., 2018).
[52] See e.g. Letter to the Editor of PNAS (17 December 2011), Spurious switching points in traded stock dynamics *at* http://www.er.ethz.ch/media/essays/PNAS.html
[53] Supported with the funds, education, involvement, and demand of the public.

costly examples of their failure, are well known. With financial markets, economies and societies more broadly, being complex and dynamic, risk and reward therein should be embedded in a prospective-evolutionary framework. In line with this, a *Darwinian (evolutionary) view* to risk models was applied (Embrechts, 2017), embracing the idea that a model that can best adapt to change (or even mutate) will win the evolutionary race. Building additionally on the concept of ecological resilience (Kovalenko and Sornette, 2013), we posit that such a class of model or process can also be thought of as best overall, even if it may be less *fit* during stable periods.

To comment on some relevant elements: Despite inadequacy in principle, the majority of risk-management activities rely on historical data and assumptions of stationarity. Scenario analysis and more general stress testing are done, and can be prospective. However, common and problematic pitfalls include inappropriate reliance on Normal distributions, linear correlations, and quantile risk-measures, which can and often have led to dangerous accumulation of residual risk. These errors have proved consequential, although the large bonuses awarded to some are not 'clawed back', and even direct losses may be socialized.

The need for going beyond single-snapshot views was emphasized by the relevant authority in Japan: "The safety and soundness of a bank cannot be captured by a point-in-time assessment" and called for moving towards a dynamic approach (Mori, 2016), echoing the proposition to transition from the static value-at-risk approach to a dynamical time-at-risk methodology (Kovalenko and Sornette, 2013). Drawing further from Embrechts: Indeed relying on single risk metrics is questionable, when the full Darwinian quality of a process is of interest; sound validation will rely on comprehensive qualitative and quantitative criteria and tests. Further, the players and regulators need to strike the right explore-exploit balance between diverse internal models and more rigid but comparable standardized models. Simplicity will be a virtue, as emphasized by Executive Director at the Bank of England, Andy Haldane (2012): "*As you do not fight fire with fire, you do not fight complexity with complexity. Because complexity generates uncertainty, not risk, it requires a regulatory response grounded in simplicity, not complexity.*" Also relevant, in the worlds of John Tukey, "*Far better an approximate answer to the right question, which is often vague, than an exact answer to the wrong question, which can always be made precise.*"

Indeed, consideration of systemic aspects of financial markets – including a fuller acknowledgment of the role of chance in markets, and a better alignment of incentives – should better support stability than the implementation of large and expensive enterprise risk management systems, based on historical data and assumptions of stationarity, creating a comfortable and convenient illusion of control.

6.7 Synthesis and review of tools and systems in the literature

The *raw outcome, risk-adjusted*, and *prospective-evolutionary* approaches are applicable in a broad range of discipline like statistics, game theory, system engineering, psychology, economics, finance, military, political and business strategy. Figure 10 gives a graphical overview (details and references in A.3). The color code is our assessment: Disciplines colored green are well-adapted to the environmental conditions. Good examples are De Mesquita's Predictioneer's game (2009), Sornette's Dragon Kings (2012), Tetlock's Superforecasters (2015) or Seligman et al.'s Homo Prospectus (2013, 2016). The red-colored examples do not account sufficiently for the absence of stationarity in the environment. A controversial example is the Capital Asset Pricing Model[54]. Other examples are the Martingale Measure, which describes a process where the most probable observation of tomorrow is the realization of today, Black and Scholes' replicating portfolio methodology (Black and Scholes, 1973) and multiple-testing/overfitting. Up-and-down economics (Krugman, 1990) is a standard practice in journalism where adaptive, non-stationary, complex systems like the economy or the financial markets are naively reported as 'went up' or 'went down', with insufficient concern for risk (or significance) adjusted and prospective nature.

---

[54] In theory, the CAPM includes all the future expectations of all agents, all of whom are supposed to optimize their portfolio and thus converge to the same 'tangent portfolio', which becomes the market portfolio. In this sense, it can be seen as prospective. In practice, it assumes equilibrium in a stationarity world and fails miserably, as witnessed by the hundreds of factors that have been proposed to complement / cure it (Harvey et al., 2016). That is why we classify it as a risk-adjusted outcome based approach, i.e. retrospective, because it extrapolates that the statistical properties of the future will be like the past.

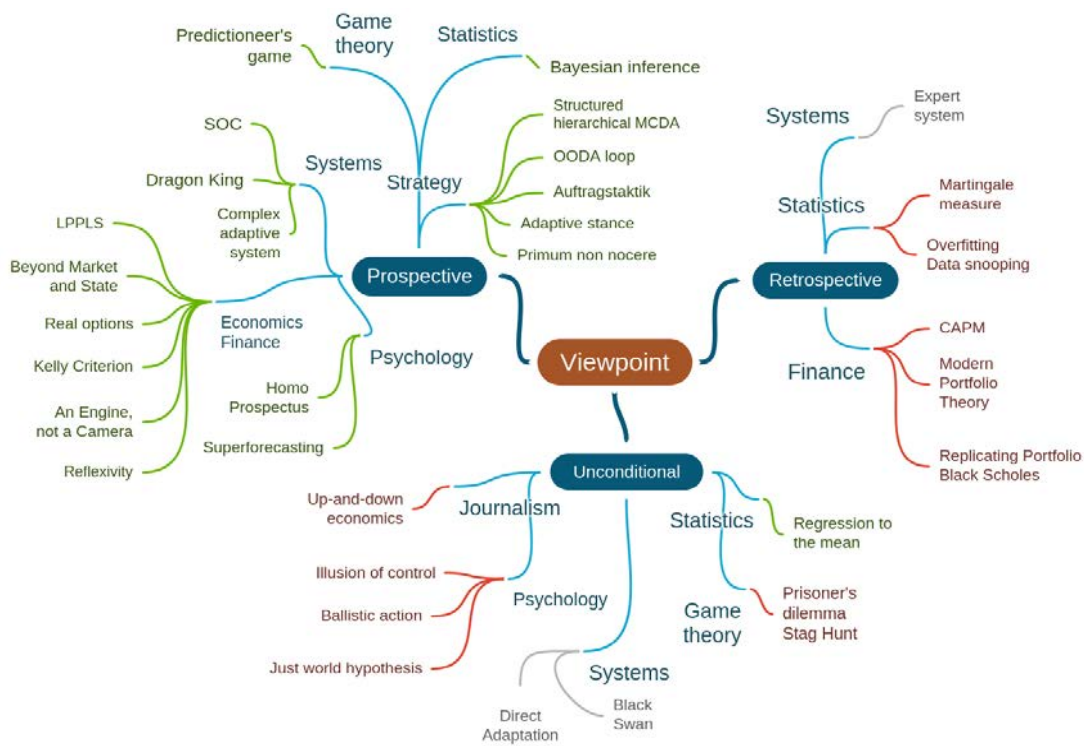

*Figure 10. A multi-disciplinary view of tools and systems from the literature classified according to viewpoint. The green-colored are fully aligned with the environment, the red-colored do not take insufficient stationarity into account, the grey-colored are just tools. More information and references are given in Appendix A.3.*

7. Concluding remarks

Thinking about fortune and the role of chance events is as old as man itself. Providing a modern update in that tradition, we noted that, nowadays, we often reward luck over skill, and in some cases are not risk-taking enough. To support a prosperous meritocratic society, we posed the *fair reward problem*, and provided means to address it.

A static framing of the top level *skill-success gap* problem, arising from the difference between the observed distribution of wealth and unobserved one of merit was introduced. The pure luck explanation was excluded by recognizing the large variation in productivity is driven by multifaceted complex tasks. The pure meritocracy edge case was excluded by detailing six important mechanisms, i.e., *wheels of fortune,* generating stochasticity in life and success.

Careful domain-specific analysis is needed to identify the role of luck in success. Taking a systems approach, depending on observability and stationarity of the system, three measures for assessing merit were defined: raw outcome (the default), risk-adjusted, and *prospective*. Performance of the first two were characterized in simulation studies, showing that risk-adjusted measures more rapidly 'separate the wheat from the chaff' than simply 'waiting for the cream to rise', under raw-outcome assessment. The third measure stands out from the others, assimilating our evolutionary past, and providing an encompassing framework for assessing real world non-stationary systems. An exemplary evolutionary process was defined, comprising five processes and comprehending exploration-exploitation cycles to refine and then overcome exhausted paradigms. Indeed, dynamics are important as the need to mutate/transform may emerge; and internal and external factors must be accounted for.

Through these lenses, a range of important applications were discussed. Selected findings include: At a large scale, we considered the political system under the *prospective view* -- e.g., with modern nation state as the current paradigm within a broader evolutionary process -- confirming the full richness of our evolutionary framework. Differentiated progressive tax rates were suggested, with higher taxes being applied to higher luck/risk fields, especially where losses are socialized. The *Amy & Frank* algorithm was introduced, to recover some wisdom of crowds in the increasingly interconnected cybernetic world, where success of viral content has little to do with merit, and ranking and aggregation of information by quality is severely inefficient. Further, science and academia were addressed, with market mechanisms identified as promising for meritocracy as well as performance in development of knowledge.

**Acknowledgements**: We are grateful to Heinrich Nax and Paul Embrechts for stimulating feedbacks.

# Appendix:
# 'Should we reward process rather than success?'

A.1 Simulating the role of time in separating skill and luck in the 'outcome based' process.

We start from a heterogenous population of N agents, each one has an individual share of skill $\mu_i$ and luck $\sigma_i$. The agents participate in an activity, the outcome of which follows a GBM process. After a time-step $\Delta T$, the success of an agent can be calculated from equation (3). As explained in section 5.2, this success has a skill part, equal to $(\mu - \frac{\sigma^2}{2}) \Delta T$, but also a luck part proportional to $\sigma_i \sqrt{\Delta T}$. Skill and luck are orthogonal, non-correlated by design. In line with Shockley's findings, we sample both $\mu_i$ and $\sigma_i$ from lognormal distributions[55]. As such, the whole population of agents is characterized by four parameters:
- $\mu_{skill}$, the mean of the skills of the N agents;
- $\sigma_{skill}$, the standard deviation of the skills of the N agents;
- $\mu_{luck}$, the mean of the luck of the N agents;
- $\sigma_{luck}$, the standard deviation of the luck of the N agents.

We sample N agents from this population, so that to each one is given a specific amount of skill ($\mu_i$) and luck ($\sigma_i$). Then, we let them act over a period $\Delta T$, during which the evolution of their success follows a GBM process of the form (3), with the idiosyncratic $\mu_i$ and $\sigma_i$. Finally, we split up the population in deciles, each containing N/10 agents, according to their realized individual success after the time-step $\Delta T$, and not using the "real" $\mu_i$ and $\sigma_i$ values of each of the agents, which are assumed not directly observable. We calculate the average of the skills $\mu_i$ and the luck $\sigma_i$ for each decile. The purpose of this simulation is to see under what conditions it is possible to separate "real" skill from "real" luck, based on one single observation of success, so as to simulate when outcome based reward can be used effectively.

The simulation is repeated for different periods $\Delta T$, which we will call the "vetting period", the period to assess the qualities of, or to "vet" the agent. The results are presented in figure A.1, for four different populations with increasing heterogeneity. As explained, each population is fully characterized by four different parameters, which are given in the uppermost subplot. The two subplots below give results for that same population. Here, the parameters are not repeated in order not to make the figures too busy.

Each subplot shows the results for eight different vetting periods: one day (1D), one week (1W), one month (1M), one quarter (1Q), half-a-year year (1H), one year (1Y), two years (2Y) and four years (4Y). In that respect, the four parameters that characterize the populations should be regarded as annualized numbers in the GBM process. The x-axis, which ranges from 1 to 10, differentiates between the different deciles, 1 is the most successful and 10 is the least successful decile after the vetting period. The y-axes give the average skill, luck and Sharpe ratio[56] for the N/10 agents in each decile. The black dotted line shows the out-of-sample value for that population, meaning, independent of decile or vetting period.

---

[55] Like the productivity in equation (1), we also see skill as a multiplication of different factors such as talent, perseverance, skill, analytical intelligence, communication abilities, charisma, emotional intelligence, social skills ... If each of these individual factors is normally distributed, combination by multiplication will result in an aggregate that is lognormally distributed.

[56] Note that the skill per decile is defined as the arithmetic mean of the $\mu_i$ of the agents in the decile, the luck is the square root of the arithmetic mean of the $\sigma_i^2$ of the agents in the decile, the Sharpe ratio is the skill per decile divided by the luck per decile. The Sharpe ratio is a standard measure of quality of financial investments, in the sense that it provides a theoretically founded risk-adjusted measure of performance.

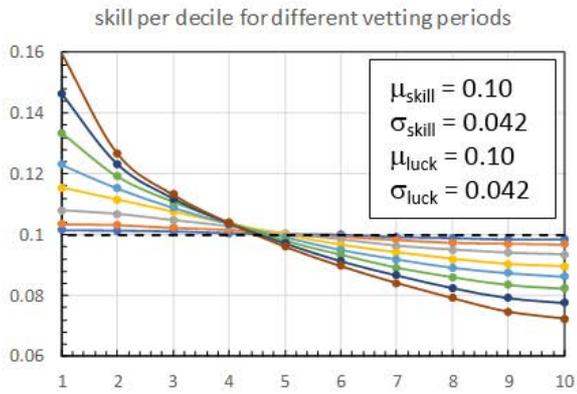
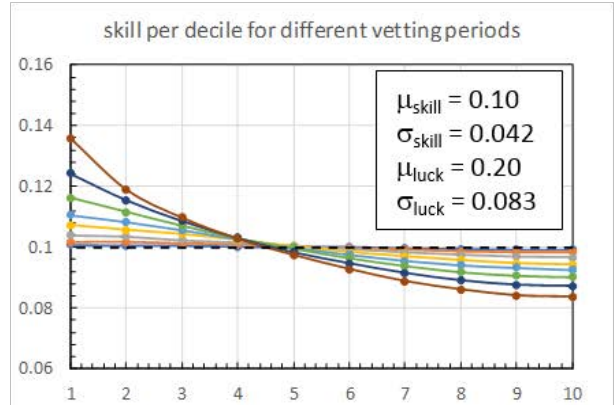
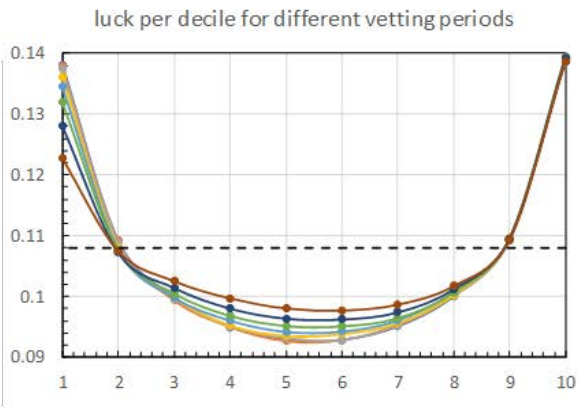
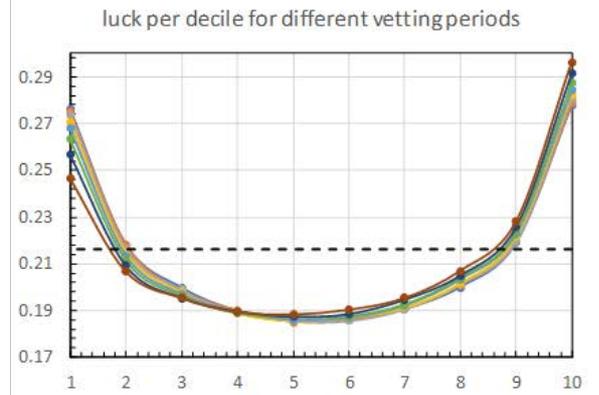
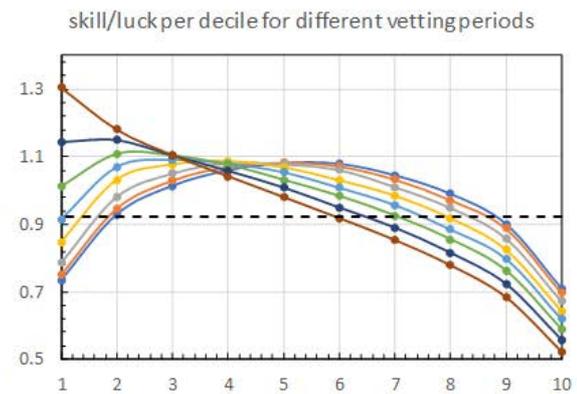
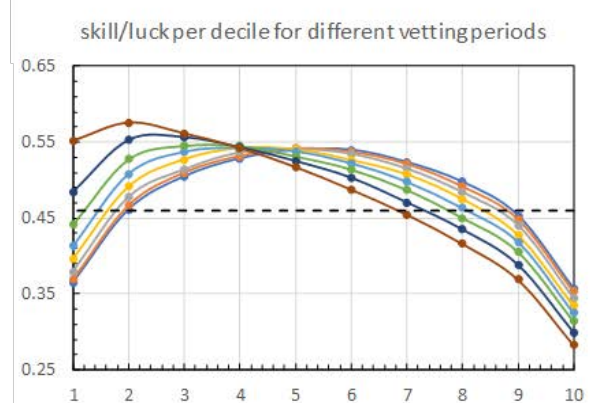

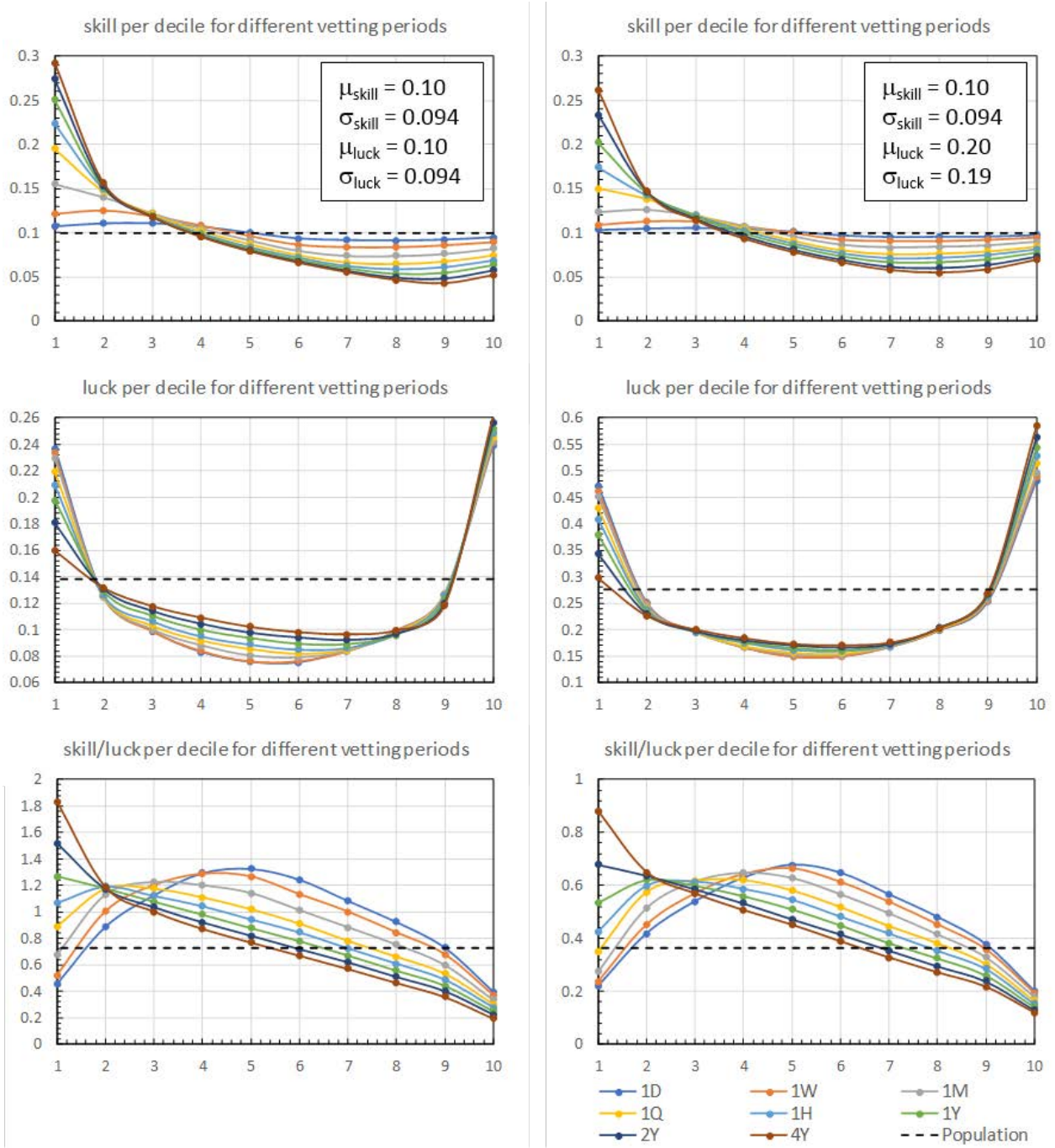

Fig. A.1: Skill, luck and Sharpe ratio (defined as skill divided by luck) per decile over one year after different vetting periods ΔT (1 day, 1 week, 1 month, 1 quarter, half-a-year, one year, two years and four years) and for two different populations, one for each column. The black dotted line shows the out-of-sample value for the whole population. To make the figure less crowded, the legend is only added to the upper subplot. The two subplots below this upper plot have the same population characteristics.

A.2 Simulating the effect of intermediate observations

We follow the same basic model as in Appendix A.1, but, instead of using the direct outcome of the process as the measure for success, now we divide this by the estimated realized volatility. This Sharpe ratio will be our new proxy for success.

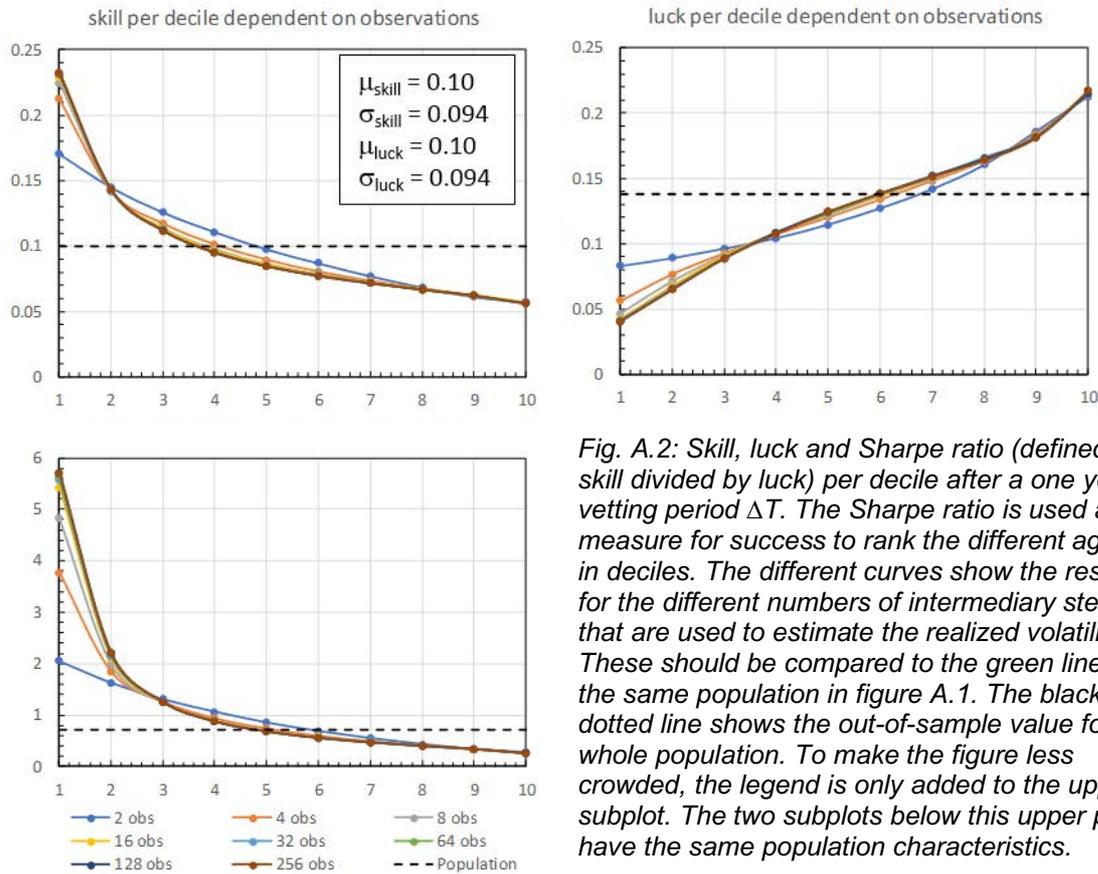

Fig. A.2: Skill, luck and Sharpe ratio (defined as skill divided by luck) per decile after a one year vetting period $\Delta T$. The Sharpe ratio is used as the measure for success to rank the different agents in deciles. The different curves show the results for the different numbers of intermediary steps that are used to estimate the realized volatility. These should be compared to the green line for the same population in figure A.1. The black dotted line shows the out-of-sample value for the whole population. To make the figure less crowded, the legend is only added to the upper subplot. The two subplots below this upper plot have the same population characteristics.

Again, we split up the population in deciles, each containing N/10 agents, according to their realized individual Sharpe ratio after the time-step $\Delta T$, and we calculate the average of the skills $\mu_i$ and the luck $\sigma_i$ for each decile. The results for an increasing number of intermediary steps, and for a vetting period of one year, are presented in figure A.2. These can be compared to the green line for the same population in figure A.1.

## A.3 Tools and systems in the literature

| Concept | Field | About | Viewpoint | Reference |
|---|---|---|---|---|
| Direct adaptation | System engineering | Control system reacting when a threshold is crossed. like a thermostat: 'too hot' ... 'too cold' ... 'too hot' ... | Unconditional | |
| Black Swan | System engineering | Exogenous unpredictable disturbance of large impact. | Unconditional | Taleb (2010) |
| Prisoner's dilemma/Stag hunt | Game theory | Tragedy of the commons/social dilemma: Nash equilibrium based on individual perspective and non-cooperation even though collaboration would end in a better outcome for every agent. | Unconditional | |
| Up-and-down economics | Journalism | Coverage of financial markets and economy: 'went up', 'went down'. | Unconditional | Krugman (1990) |
| Illusion of Control | Psychology | Estimation of personal success probability inappropriately higher than the objective probability. | Unconditional | Langer (1975) |
| Just world hypothesis | Psychology | Good things happen to people who do good things, bad things happen to people who do bad things. | Unconditional | Lerner and Montada (1998) |
| Ballistic action | Psychology | Logic of Failure: Taking measures without checking the effect of these measues later. | Unconditional | Dorner et al. (1990) |
| Expert system | Control theory | Expert control system based on a decision tree, 'if ... then, ... else. | Retrospective | |
| CAPM | Finance | Capital Asset Pricing Model: pricing financial assets based on 'tangent portfolio'. | Retrospective | Sharpe (1964) |
| Replicating portfolio | Finance | Derivative pricing based on dynamic replication and efficient markets, fundamental principal behind the Black-Scholes formula for option pricing. | Retrospective | Black and Scholes (1973) |
| Modern Portfolio Theory | Finance | Mean variance analysis, mathematical framework for structuring a porfolio of financial asets accounting for diversification. | Retrospective | Markowitz (1952) |
| Overfitting | Statistics | Datasnooping: spurious correlation found as the result of extensive data analysis, without any causation. | Retrospective | |
| Martingale | Statistics | Random walk (or stochastic process) for which, at a particular time, the most probable next value, given all prior values, is equal to the present value. | Retrospective | |

*Table A.1: Literature review: tools and systems based on outcome based reward and risk-adjusted outcome based reward*

| Concept | Field | About | Viewpoint | Reference |
|---|---|---|---|---|
| Predictioneer's game | Game theory | A game theoretical framework based on an understanding of the utility function of agents and decision makers in a geopolitical context. | Prospective | De Mesquita (2009) |
| Complex adaptive system | System engineering | Self-learning control system inspired by biological evolution based on a scheme that changes based on reproduction, mutation, fitness function (genetic algorithm). | Prospective | Gell-Mann (1994) |
| Dragon King | System engineering | Endogenous predictible disturbance of large impact. These events are the outcome of a dynamical system progressively approaching an instability leading to a transition to another mode. | Prospective | Sornette and Ouillon (2012) |
| Beyond markets and states | Economics | Through communication and cooperation, institutions may self-organize beyond the market and the state. Complex systemic problems may best be solved by complex self-organized institutions. | Prospective | Ostrom (2010) |
| An Engine, not a camera | Finance | Financial and economic models are not merely descriptive but feed back to and influence reality. | Prospective | MacKenzie (2006) |
| Reflexivity | Finance | Observers are part of the system observed: feedback loop between investors' perception and the environment. | Prospective | Soros (1998) |
| LPPLS | Finance | Log Periodic Power Law Singularity: estimation of a structural break in the future, time at risk. | Prospective | Sornette (2017) |
| Kelly criterion | Finance | Optimizing the size of a bet, taking into account the gambler's edge or expectation of the odds. | Prospective | Kelly Jr (1956) |
| Real options | Finance | Take into account future potential when valuing financial assets. | Prospective | Dixit and Pindyck (1994) |
| Primum non nocere | Medicine | Hippocratic oath: First do no harm, typified by the Yellowstone effect. | Prospective | Malamud et al. (1998) |
| Structured Hierarchical MCDA | Strategy | Structured Hierarchical Multi Criteria Decision Analysis: Evaluation of multiple conflicting, hierarchical criteria in decision making. | Prospective | Mearns and Sornette (2018) |
| Adaptive stance | Strategy | Counters predispositions and offers an effective methodological framework for managing, creating, shaping and interacting with complex systems. | Prospective | Grisogono and Radenovic (2011) |
| Auftragstaktik | Strategy | Mission command tactics, global mission carried out by local decisions. Strategic warfare based on Moltke, Klausewitz, Sun Tzu's Art of War, Napoleon | Prospective | |
| OODA loop | Strategy | Observe, orient, decide, act, use time as a weapon, be the fastest to complete the cycle from observation to action, act on information faster than the opponent. | Prospective | Coram (2004) |
| Lean enterprise | Strategy | The company who acts on information faster is in the best position to win, business strategy based on John Boyd's OODA loop | Prospective | Humble et al. (2015) |
| Homo Prospectus | Psychology | 'Forward Looking Man' for whom consciousness is the generation of simulations about possible futures. | Prospective | Seligman et al. (2013, 2016) |
| Superforecasting | Psychology | Organizing forecasting tournaments to identify the distinctive features of good quality expert judgements and predictions. | Prospective | Tetlock and Gardner (2015) |
| Self-organized criticality | Statistical physics | A dynamical system that tunes itself as it evolves towards criticality, well-known example is a sandpile. | Prospective | Bak et al. (1987) |
| Regression to the mean | Statistics | After a succesful series of statistical independent observations (a hot streak), there is a high probability that success wanes and observations reverse to the mean. | Prospective | Galton (1886) |
| Bayesian inference | Statistics | Use of Bayes' theorem to update probability for a hypothesis in real time while more information becomes available. | Prospective | |

*Table A.2: Literature review: tools and systems based on prospective reward.*